\documentclass[11pt]{article}
 \pdfoutput=1
\usepackage{jheppub}

\usepackage{color}
\usepackage{amsmath}
\usepackage{verbatim}
\usepackage{subfigure}
\usepackage{acronym}
\usepackage{amsfonts}
\usepackage{amssymb}
\usepackage{mathrsfs}
\usepackage{graphicx}
\usepackage{multirow}
 \usepackage{slashed}
 \usepackage{epsfig,multicol,bbm}
 \usepackage{url}

\def\beq{\begin{equation}\begin{aligned}}
\def\eeq{\end{aligned}\end{equation}}
\newcommand{\kahler}{K\"{a}hler }
\def\OO{\mathcal{O}}

\title{Precision Unification and Proton Decay in F-Theory GUTs with High Scale Supersymmetry}

\date{}

\author[1]{Arthur Hebecker}  
\author[2]{and James Unwin}
\affiliation[1]{Institut f\"ur Theoretische Physik, Universit\"at Heidelberg, Philosophenweg 19, Heidelberg, Germany}
\affiliation[2]{Department of Physics, University of Notre Dame, Notre Dame
IN, 46556, USA}
\emailAdd{A.Hebecker@ThPhys.Uni-Heidelberg.de}
\emailAdd{james.unwin@nd.edu}

\abstract{
F-theory GUTs provide a promising UV completion for models with approximate gauge coupling unification, such as the (non-supersymmetric) Standard Model. More specifically, if the superparters have masses well above the TeV scale, the resulting imperfection in unification can be accounted for by the, in principle calculable, classical F-theory correction at the high scale. In this paper we argue for the correct form of the F-theory corrections to unification, including KK mode loop effects. However, the price of compensating the imprecise unification in such High Scale SUSY models with F-theory corrections is that the GUT scale is lowered, potentially leading to a dangerously high proton decay rate from dimension-6 operators.  We analyse the possibility of suppressing the decay rate by the localization of $X,Y$ gauge bosons in higher dimensions. While this effect can be very strong for the zero modes, we
find that  in the simplest models of this type it is difficult to realize a significant suppression for higher modes (Landau levels).  Notably, in the absence of substantial suppressions to the proton decay rate, the superpartners must be lighter than 100 TeV to satisfy proton decay constraints. We highlight that multiple correlated signals of proton decay could verify this scenario.
}

\begin{document}

\hfill \vspace{-5mm} May 12, 2014

\maketitle


\section{Introduction}

The lack of supersymmetry at the LHC, or indeed of any sign of `New Physics', has led to a renewed interest in models which abandon the principle of naturalness and suppose that there is an inherent fine tuning due to environmental selection in the Higgs sector, see e.g.~\cite{ArkaniHamed:2004fb,Unwin:2011ag,Giudice:2004tc,Wells:2003tf,Chatzistavrakidis:2012bb,Arvanitaki:2012ps,Hall:2009nd,ArkaniHamed:2006mb,ArkaniHamed:2012gw,Unwin:2012fj,Hall:2011jd,Hebecker:2012qp,Hebecker:2013lha,Ibanez:2012zg,  Giudice:2011cg,Baryakhtar:2013wy,Ibanez:2013gf,Hisano:2012wm,Hisano:2013cqa,Hisano:2013exa,Hall:2013eko,Hall:2014vga,Liu:2012qua}. With technical naturalness no longer imposed, alternative guiding principles in the search for theories `beyond the Standard Model' (BSM) are required.
It is reasonable to suppose that 4d supersymmetry (SUSY) exists at some scale as it is a necessary ingredient in all quantitatively controlled string-theoretic UV completions. Furthermore, as we will argue in Sect.~\ref{S2}, the GUT picture continues to provide a strong guiding principle for BSM model building even in the absence of TeV-scale SUSY.
Motivated by this, we examine the phenomenology of F-theory gauge coupling unification \cite{Donagi:2008ca,Beasley:2008dc,Donagi:2008kj,Donagi:2008kj2,Blumenhagen:2008zz,Blumenhagen:2008aw} in models of High Scale SUSY\footnote{We shall use `High Scale SUSY' to mean that the non-SM states have masses well above the TeV scale. This includes models which some authors call  `Intermediate Scale SUSY'.}  \cite{Hall:2009nd,Giudice:2011cg,Unwin:2011ag,Hebecker:2012qp,Ibanez:2012zg, Liu:2012qua,Hisano:2012wm,Hall:2011jd,Unwin:2012fj,Ibanez:2013gf,Hisano:2013exa,Hisano:2013cqa, Hebecker:2013lha,Hall:2013eko}, following the work of Ibanez, Marchesano, Regalado and Valenzuela \cite{Ibanez:2012zg}.

As is well known, gauge coupling unification in models where all superpartner masses are far above the TeV domain is wanting. However, it is quite plausible that the loss of precision unification can be attributed to large (often unknown) UV threshold corrections.  In particular, F-theory GUTs feature well-defined UV corrections of this type, which can be estimated and may account for the apparent loss of precision in gauge coupling unification. In Sect.~\ref{S3} we review the literature of F-theory threshold effects to gauge coupling unification and argue that it is consistent and necessary to include both the classical part from the DBI action \cite{Blumenhagen:2008aw,Mayrhofer:2013ara,Donagi:2008kj} and the KK loop effect \cite{Donagi:2008kj2}. Subsequently, we study the parametric dependence of the (F-theory) GUT scale on the SUSY scale.  In contrast to \cite{Ibanez:2012zg}, we allow for the flux superpotential, and hence $m_{3/2}$, to be parametrically small (e.g.~through tuning of fluxes) such that the SUSY scale can vary (together with the magnitude of the F-theory threshold corrections \cite{Mayrhofer:2013ara}) between very high and moderately low values.  We also study, in Appendix \ref{S32}, the running and F-theory unification of gauge couplings in models with multiple mass scales (relevant e.g.~for situations with a strongly split superpartner spectrum).

As the superpartner mass scale is raised the GUT scale is lowered, thus, the null-observation of proton decay might be the most significant problem facing GUTs of this type.  Consequently, it is interesting to look for well-motivated mechanisms in which the proton decay rate is suppressed, see e.g.~\cite{Hall:2001pg,Dorsner:2004xa,Hebecker:2002rc}.  It has been suggested that this can be achieved in higher-dimensional models due to localisation of $X,Y$ gauge bosons \cite{Ibanez:2012zg,Hamada:2012wj,Kakizaki:2013ila}.  
In Sect.~\ref{S3b} we analyse proton decay suppression in F-theory GUTs using toy models. The simplest and strongest suppression mechanism arises if, due to gauge flux, the $X,Y$ bosons are localized away from the matter curve. We show that this localization effect is undone by the contributions of higher Landau levels of the $X,Y$ gauge fields. The latter unavoidably spread out and can not be separated from the SM matter fields. Thus, at least in the simplest scenarios, the {\em a priori} possible exponential suppression of proton decay can not be achieved. While this does not rule out a weaker, non-exponential suppression, we take our findings as a motivation to consider the constraints on F-SU(5) High Scale SUSY under the assumption that $X,Y$-induced proton decay proceeds with the same rate as in 4d GUTs. This analysis is performed in Sect.~\ref{S4} and leads to strong constraints on the spectrum.
 We find that in order to satisfy experimental constraints on proton decay, superpartners must have sub-100 TeV masses.  Additionally, we comment on the possibility of observing signatures of High Scale SUSY at next-generation proton decay experiments. Notably, one might expect correlated signals of proton decay in $\pi^0e^+$, and $K^+\bar{\nu}$, the first due to exchange of $X,Y$ bosons and the latter mediated by GUT scale triplet Higgses.


\section{Motivating GUTs without weak scale SUSY}
\label{S2}

The successful prediction of the weak mixing angle from SU(5) boundary conditions \cite{Dimopoulos:1981zb} is a notable experimental hint that unification might be realised in nature. Moreover, from a theoretical perspective, GUTs provide a framework for understanding the SM matter spectrum. 
One might argue that the particle content of the SM can be determined solely by the requirement that the gauge theory is anomaly free. The fact that it fills out representations of SU(5) is then to some extent accidental. Indeed, anomaly cancellation is a strong constraint on the particle spectrum which is automatically satisfied in consistent string theory compactifications. The SM could thus arise from three or more intersecting brane stacks, without the need for unification into a larger simple group \cite{Blumenhagen:2005mu,Cremades:2002cs,Blumenhagen:2003jy}.  

Given the SM gauge group,\footnote{
We shall assume that the SM gauge group is the (minimal) necessary gauge structure required for anthropic reasons, however there is some ongoing debate regarding this assumption \cite{Harnik:2006vj,Clavelli:2006di,Gedalia:2010iy,Gato-Rivera:2014afa}.} 
it is interesting to identify the minimal anomaly-free, chiral spectra containing the bi-fundamental representation $\big(3,2\big)_Y$. Here $Y$ is the (so far unspecified) hypercharge. There are three gauge anomaly cancellation conditions as well as a mixed gauge-gravity condition:
\beq
{\rm SU(2)^2U(1)}:& \quad  \sum_{\square~{\rm of~SU(2)}}  Y_i =0 ~,
\hspace{12mm}
{\rm U(1)^3}: \quad  \sum_{\rm all~repr.} Y_i^3 =0 ~,\\[5pt]
{\rm SU(3)^2U(1)}:& \quad  \sum_{\square~{\rm of~SU(3)}} Y_i =0 ~,
\hspace{9mm}
{\rm G^2U(1)}: \quad   \sum_{\rm all~repr.} Y_i =0~.
\eeq
Simultaneously satisfying these four conditions strongly restricts the combinations of representations which are allowed and only the following (minimal) spectra, given in terms of left-handed spinors, satisfy these requirements \cite{Foot:1988qx,Knochel:2011ng}
\begin{equation*}
\begin{aligned}
{\bf I:}\quad&(3,2)_{1/6}\oplus(\overline{3},1)_{-2/3}~\oplus(\overline{3},1)_{1/3}~\oplus(1,2)_{-1/2}~\oplus(1,1)_{1}\\[5pt]
{\bf  II:}\quad&(3,2)_Y~~\oplus(\overline{3},1)_{-Y-1/2}~\oplus(\overline{3},1)_{-Y+1/2}~\oplus(1,2)_{-3Y}~\oplus(1,1)_{3Y-1/2}~\oplus(1,1)_{3Y+1/2}\\[5pt]
{\bf  III:}\quad&(3,2)_Y~~\oplus(\overline{3},2)_{-Y-1/2}~\oplus(\overline{3},2)_{-Y+1/2}~\oplus(3,2)_{-Y}~\oplus(\overline{3},2)_{Y-1/2}~\oplus(\overline{3},2)_{Y+1/2}\,.
\end{aligned}
\end{equation*}

Case $\bf I$ corresponds to a single SM generation or, equivalently, the decomposition of $\bf \bar 5$ and $\bf 10$ of SU(5) in terms of the SM group. Moreover, it features the smallest field content of the three cases. To account for right-handed neutrinos, it could be supplemented by a $(1,1)_{0}$. The sets of representations {\bf II} and {\bf III} possess a free parameter $Y$. The fact that the hypercharge is no longer quantised can be understood as follows: Sets {\bf II} and {\bf III}, viewed as SU(3)$\times$SU(2) representations, remain anomaly-free for two {\it different} U(1) charge assignments. The parameter $Y$ specifies which linear combination of the two we choose to gauge. Indeed, {\bf II} can be viewed as a SM generation plus right-handed neutrino, with both SM-hypercharge and $B\!-\!L$ anomaly free.  The choice $Y=1/6$ corresponds to gauging hypercharge. In this case $\bf II$ reduces to $\bf I$ supplemented with a $(1,1)_{0}$. Even with $Y=1/6$, set $\bf II$ still has two anomaly-free U(1) factors. However, the subtraction of $(1,1)_{0}$ causes $B\!-\!L$ to become anomalous. Hence {\bf I} (the SM) is also distinguished as the case with smallest symmetry structure.

Thus, if the matter content of our universe is `selected' to be the smallest anomaly free set of representations which allows communication between the various gauge symmetries, then this would pick-out the SM. In addition, the smallness of the symmetry group selects {\bf I} over {\bf II} and {\bf III}. 
Finally, it is easy to accept that we live in case {\bf I} rather than {\bf II} or {\bf III} by pure accident. As a result, the motivation for GUTs as an organizing principle appears to be lost.
However, a puzzle remains: If the particular choice of anomaly free chiral matter is accidental, then it is surprising that it is replicated thrice in the generation structure of the SM. Instead, some combination of the different anomaly free choices presented above appears more likely. On the contrary, 
the replicated structure appears automatically in the framework of SU(5) and SO(10) GUTs. Restricting the field content of the GUT to anomaly free combinations of low dimension representations (as is the expectation from string theory) automatically requires that generations of low energy field content be identical.\footnote{At the more technical level, the replication is a natural outcome of an appropriate flux choice.} We view this as one of the great successes of GUT models which could survive a non-discovery of TeV-scale superpartners. Based on this, we believe that the GUT idea remains a prominent motivation for BSM model building.


\section{F-theory unification in High Scale SUSY}
\label{S3}

It is non-trivial to UV complete GUTs into string theory; in many instances problems arise such as incorrect Yukawa structure or inappropriate field content. One particularly promising possibility is found in the framework of F-theory. In F-theory GUTs, SM matter resides on 2D sub-manifolds, {\em matter curves}, within a 4d manifold, the {\em GUT brane}, on which SU(5) gauge fields are localised. All of this is embedded in the 6d compactification space. Couplings between states are determined by the overlap of their associated wavefunctions, which describe the localisation in the compact dimensions \cite{Klebanov:2003my,Camara:2011nj,Camara:2013fta,Hayashi:2009bt,Palti:2012aa,Heckman:2008qa,Conlon:2009qq, Font:2009gq, Beasley:2008kw,Cecotti:2009zf,Aparicio:2011jx,Font:2012wq,Font:2013ida,Hamada:2012wj}.

\subsection{String theory relations involving GUT parameters}

In this section we provide a careful discussion of the connection between the GUT scale, the GUT coupling, and the SUSY breaking scale. We follow primarily the conventions of  \cite{Giddings:2001yu}.  Suppose, for simplicity, that the GUT brane is a torus with volume
\beq
{\cal V}_{\rm GUT}^s=(2\pi R/l_s)^4~,
\label{qq}
\eeq
with the string length $l_s\equiv2\pi\sqrt{\alpha'}$. We identify the Kaluza-Klein (KK) and GUT scale, $M_{\rm KK}\equiv M_{\rm GUT} = 1/R$. Recalling the prefactor of the 10d Einstein-Hilbert term in the string frame, $1/2\kappa_{10}^2g_s^2=2\pi/l_s^8g_s^2$, the GUT and Planck scale are then related by
\beq
\frac{M_{\rm GUT}^2}{\overline{M}_P^2}=\frac{R^{-2}}{4\pi(g_sl_s)^{-2}
{\cal V}^s}\,,
\label{ccc}
\eeq
where ${\cal V}^s$ is the string frame volume of the Calabi-Yau, in units of $l_s^6$. Translating from the string-frame to the Einstein-frame, ${\cal V}^s=g_s^{3/2}{\cal V}$ and ${\cal V}_{\rm GUT}^s=g_s{\cal V}_{\rm GUT}$, and returning to 4d supergravity conventions with $\overline{M}_P^2=M_P^2/(8\pi)=1$ gives the relation
\beq
M_{\rm GUT}=\frac{\sqrt{\pi}}{{\cal V}_{\rm GUT}^{1/4}{\cal V}^{1/2}}\,.
\label{bb}
\eeq
Further, since the volume of the GUT 4-cycle sets the SU(5) gauge coupling at the GUT scale
\beq
\alpha_{\rm GUT}^{-1}=2{\cal V}_{\rm GUT}\,,
\label{bbb}
\eeq
it follows that
\beq
M_{\rm GUT}=(2\alpha_{\rm GUT})^{1/4}\sqrt{\pi/\mathcal{V}}~.\label{mgut}
\eeq
Additionally, we shall use that in supergravity the soft masses generically depend on $|W_0|$ as\footnote{
We
use K\"ahler and superpotential normalization of \cite{Giddings:2001yu}, which implies that $W_0$ is a sum of products of integer flux numbers and complex-structure periods, i.e.~it is naturally an ${\cal O}(1)$ quantity.}
\beq
M_{\rm SUSY}\simeq m_{3/2}=\frac{\sqrt{g_s}|W_0|}{4\sqrt{\pi}\mathcal{V}}~.
\eeq
Eliminating ${\cal V}$ with the help of eq.~(\ref{mgut}) yields 
\beq
M_{\rm SUSY}\simeq \frac{g_s^{1/2}|W_0|}{2(2\pi)^{3/2}}\,\frac{M_{\rm GUT}^2}{\alpha_{\rm GUT}^{1/2}}\,.
\label{alp}
\eeq
According to \cite{Ibanez:2012zg}, this can be used to fix the (high) SUSY breaking scale as follows: One sets $g_s$ and $W_0$ to their natural ${\cal O}(1)$ values, preserves precision unification using F-theoretic GUT-scale corrections, and supplements eq.~(\ref{alp}) with the appropriate RGEs linking GUT and SUSY breaking scale. Ignoring also the $2\pi$ factors, this gives $M_{\rm GUT}\simeq3\times10^{14}$ GeV and $M_{\rm SUSY}\simeq5\times10^{10}$ GeV \cite{Ibanez:2012zg}. 

We take a different perspective: $W_0$ can easily be tuned to be much smaller than ${\cal O}(1)$ by the choice of fluxes. Then the above logic does not fix $M_{\rm SUSY}$, leaving it as a free parameter linked to $|W_0|$. 
Our investigation of proton decay suggests that, at least in the arguably most natural models, it is difficult to substantially suppress dimension-6 proton decay. If this were true more generally, the above scenario with $|W_0|\sim g_s \sim 1$ would then be ruled out and we would actually be forced to tune $W_0$ to reduce the SUSY scale. In subsequent sections we shall explore (and constrain) the full parameter space of these F-SU(5) High Scale SUSY models. In particular we shall argue that this typically requires a sub-100 TeV SUSY scale.\footnote{Such intermediate SUSY scales may arise in certain string constructions, see e.g.~\cite{Antoniadis:2004dt,Bizet:2014uua}.}

\subsection{The form of F-theory threshold corrections}

In F-theory models the GUT group is normally broken by hypercharge flux  \cite{Donagi:2008kj2,Beasley:2008kw}, a non-trivial field strength proportional to the hypercharge generator along the compact dimensions, 
which leads to well-defined UV threshold corrections to the RGEs of the SM gauge couplings.
Since the quantitative results appearing in the literature \cite{Blumenhagen:2008aw,Mayrhofer:2013ara,Dolan:2011aq,Donagi:2008kj,Donagi:2008kj2}  are not (at least not obviously) consistent, and as the differences are crucial for our purposes, we begin by briefly reviewing the situation and explaining our point of view. 

The dominant high-scale corrections to gauge coupling unification arise from (the parity-even part of) the correction to the Tr$\left[F^2\right]$ non-abelian 8d gauge theory action, coming from the quartic term Tr$\left[F^4\right]$. The contracted quartic field strength is defined as follows
\beq
{\rm Tr}\left[F^4\right] =
16~{\rm Tr}\big[&F^{\mu\nu}F_{\rho\nu}F_{\mu\tau}F^{\rho\tau}
+\frac{1}{2}F^{\mu\nu}F_{\mu\tau}F^{\rho\tau}F_{\rho\nu}\\
&~-\frac{1}{4}F^{\mu\nu}F_{\mu\nu}F^{\rho\tau}F_{\rho\tau}
+\frac{1}{8}F^{\mu\nu}F^{\rho\tau}F_{\mu\nu}F_{\rho\tau}\big]~.
\eeq 
This form comes from contraction of the quartic field strength ${\rm Tr}\left[F_{\mu_1\nu_1}F_{\mu_2\nu_2}F_{\mu_3\nu_3}F_{\mu_4\nu_4}\right]$ with the 10d extension of the 8d light cone gauge `zero mode' tensor, see e.g.~\cite{Tseytlin:1995bi}.
Traces in adjoint ${\rm Tr}_{\rm Adj}[~\cdot~]$ and fundamental ${\rm Tr}_{f}[~\cdot~]$ representations can be related via \cite{Metsaev}
\beq
{\rm Tr}_{\rm Adj}\left[F^4\right] = (N+8l)~{\rm Tr}_{f}\left[F^4\right]+3~{\rm Tr}_{f}\left[F^2\right]^2~,
\label{tr0}
\eeq
with $l=-1,0$ for an SO(N) or U(N) gauge theory, respectively.

From the perspective of (perturbative) type IIB theory \cite{Blumenhagen:2008aw,Mayrhofer:2013ara}, the crucial quartic term arises at tree level in the expansion of the non-abelian DBI action,\footnote{
It 
is actually more convenient to derive the term starting from the Chern-Simons action and appealing to supersymmetry, but this is a mere technicality (cf.~\cite{Blumenhagen:2008aw}).
}
 which symbolically reads
\beq
S\sim \frac{1}{g_s}\,\int\,d^8x\, \mbox{Tr}_{f}\left\{[-\mbox{det}(g_{\mu\nu}+F_{\mu\nu})]^{1/2}\right\}\,.
\eeq
In addition, the 8d gauge theory generates this quartic term at one-loop \cite{Bachas:1996bp,Green:1982sw,Metsaev} dressed by a factor ${\rm Log}(\Lambda/m)$ involving a UV cut-off $\Lambda$, at which uncancelled tadpoles in the local model are resolved \cite{Donagi:2008kj2}, and the physical mass scale $m$ required to cut-off the IR divergence (i.e.~the KK scale). Therefore, one expects the full quartic, 8d Lagrangian to be of the form
\beq
{\mathcal{L}}\sim\frac{1}{g_s}\mbox{Tr}_{f}\left[F^4\right]+\mbox{Tr}_{\rm Adj}\left[F^4\right]{\rm Log}(\Lambda/m)\,.
\label{r0}
\eeq

On the other hand, one can argue on the basis of the dual heterotic description \cite{Donagi:2008kj2}. To do so, let us think of our 8d gauge theory as arising from the type I string in 10d, compactified on $T^2$. Via two T-dualities, this corresponds to type IIB theory on $T^2/\mathbb{Z}_2$ with 32 D7 branes on top of one of the four O-planes. Further, this is dual to the SO(32) heterotic string on $T^2$. Clearly, we can switch on Wilson lines (or equivalently displace branes) to break to a desired smaller gauge group. For the group SO(32) the trace identify eq.~(\ref{tr0}) reduces to
\beq
{\rm Tr}_{\rm Adj}\left[F^4\right] = 24\left({\rm Tr}_{f}\left[F^4\right]+\frac{1}{8}~{\rm Tr}_{f}\left[F^2\right]^2\right)~.
\label{tr}
\eeq
As is well-known the type I tree-level $F^4$-term receives no loop correction in 10d \cite{Bachas:1996bp,Tseytlin:1995fy,Tseytlin:1995bi}. By contrast, the 10d heterotic string has no tree-level $F^4$-Lagrangian, reproducing however the equivalent term at the one-loop level. 
Moreover, there are no corrections beyond one loop. However, after dimensional reduction (in particular upon compactifying to 8d), the situation is more involved \cite{Bachas:1996bp}: On the type I side, one now has both tree-level and higher-loop contributions as well as instantons. Conversely, on the heterotic side, one-loop exactness is believed to persist, such that all those effects are accounted for in the one-loop heterotic term \cite{Bachas:1996bp}
\beq
{\mathcal{L}}\sim&~R_I^2\left[\frac{1}{g_I}\mbox{Tr}_f\left[F^4\right]+\left\{\int_0^\infty dl \sum_{w} e^{-w^2l/2\pi}\right\}\left(\mbox{Tr}_f\left[F^4\right]+\frac{1}{8}\mbox{Tr}_f\left[F^2\right]^2\right)\right]
+\cdots~,
\label{ell}
\eeq 
where $w$ runs over the non-zero vectors of the $T^2$ lattice (the normalization is such that $w=2\pi R_I n$ for a single $S^1$). It is important to note that this integral is log-divergent. The ellipsis contains further terms which are suppressed by powers of the coupling and are associated to integrals which are finite. These terms will give parametrically smaller contributions since, once the divergent integrals are properly regularised by the cut-off, the log-divergences are converted to logarithmic enhancements. By eq.~(\ref{tr}) the above can be rewritten as
\beq
{\mathcal{L}}\sim&~R_I^2\left[\frac{1}{g_I}\mbox{Tr}_f\left[F^4\right]+\frac{1}{24}\left\{\int_0^\infty dl \sum_{w} e^{-w^2l/2\pi}\right\}{\rm Tr}_{\rm Adj}\left[F^4\right] \right]
+\cdots~.
\eeq 
This is the relevant part of the heterotic one-loop result expressed in type I variables. We now T-dualize (twice) to type IIB as explained above and rewrite this result accordingly (i.e.~we use $R_{IIB}\simeq 1/R_I$ and $g_s\equiv g_{IIB}\simeq g_I/R_I^2$):
\beq 
{\mathcal{L}}\sim\frac{1}{g_s}\mbox{Tr}_f\left[F^4\right]+\mbox{Tr}_{\rm Adj}\left[F^4\right] {\rm Log}(1/\epsilon)~.
\label{st}
\eeq
Here we have rescaled the integration variable $l$ by $R_I^2$ to make it dimensionless and cut-off the log-divergent integral by an arbitrary small number $\epsilon\ll 1$, keeping only the logarithmic term. 

It is now apparent that the structure of the type IIB result given in  eq.~(\ref{r0}), with its (DBI-action-induced) tree-level and (field-theoretic) one-loop piece is perfectly reproduced by the one-loop heterotic string result eq.~(\ref{st}). A more careful matching, including overall numerical factors,
IR and UV cutoffs and ideally even finite (non-log-enhanced) parts would be desirable, but we leave this to future work.
For our present, phenomenological purposes, we merely take this as justification for including the tree-level $F^4/g_s$ term in what follows. This approach differs from the suggestion of \cite{Donagi:2008kj2} which restricts the correction to only the loop-contribution due to an expectation that heterotic one-loop exactness is maintained in 8d.
We believe that, at least as long as no full, string theoretic one-loop calculation is feasible in the heterotic dual of a realistic type-IIB/F-theory geometry (i.e.~as long as one simply restricts oneself to the field-theoretic logarithm), it makes sense, referring to the result of \cite{Bachas:1996bp}, to include the tree level DBI-piece as well  (this is in line with \cite{Donagi:2008kj}). 

We next briefly comment on the expected size of $\Lambda$: Most naively, one would want to identify $\Lambda$ with the local string scale $M_s$ of the type IIB theory. However, it has been argued in \cite{Conlon:2009qa,Conlon:2009xf,Conlon:2009kt,Donagi:2008kj2} that the cutoff could be enhanced to the scale at which the tadpoles of the GUT brane are cancelled. In this case $\Lambda$ corresponds to the mass of a stretched or winding open string of appropriate length. Hence, the two options are $\Lambda\sim M_s{\cal V}_{\rm GUT}^{1/4}$ (local tadpole cancellation, cf.~\cite{oai:arXiv.org:hep-th/0610007}) or $\Lambda\sim M_s{\cal V}^{1/6}$ (global cancellation). Further, one might expect that all or part of such an enhanced logarithmic correction can be captured by using an appropriate local value of $g_s$ in the tree-level term. To the best of our knowledge this has not been completely clarified in the present setting. It should be possible to do so by carefully analysing the type I calculation of \cite{Bachas:1996bp},\footnote{
The result of \cite{Bachas:1996bp} contains an open-string UV divergence proportional to tr$(F^2)^2$, which the authors subtract because it signals the exchange of a massless state. After that, an open-string IR divergence is nevertheless present. In our understanding, applied to our setting, both are IR divergences of the GUT-brane theory, the first due to tree-level exchange of closed-string light states, the second due to gauge-theory loops and possibly also tree-level gauge-boson exchange. Both are cut-off in the IR by $M_{\rm KK}$.}
where the divergence is cut-off in the UV by SO(32) KK states which, after breaking the gauge group and T-dualising in the two compact dimensions, corresponds to open strings stretched between (in this case parallel) D7 branes.

In what follows, we focus on models where the tadpole is cancelled locally, such that 
\beq
M_s\lesssim \Lambda \lesssim {\cal V}_{\rm GUT}^{1/4}M_s
\label{la}
\eeq
and the logarithm can not become very large. We will see that, in this case, the tree-level $F^4/g_s$ term tends to dominate even at  $g_s\sim {\cal O}(1)$ and, thus, the somewhat uncertain value of $\Lambda/M_{\rm KK}$ (or $1/\epsilon$) does not significantly disrupt predicitivity.

The coefficients of the (purely field-theoretic) logarithmic corrections can be related to the holomorphic torsion \cite{Donagi:2008kj2}, via which they can (in principle) be calculated, this is discussed in slightly more detail in Appendix \ref{apB}.

\subsection{F-theory corrections to gauge coupling unification}

Coming now to precise formulae, we will not perform a re-derivation along the lines given above, but rather use the most recent results in the literature.
For F-theory GUTs with High Scale SUSY there are three types of corrections to the gauge coupling RGEs: There are both tree and loop level F-theory corrections, as discussed above, and additionally raising the SUSY scale above $m_Z$ affects the RGE evolution. The signs and magnitudes of these effects are such that, in principle, they can compensate each other and maintain successful TeV-MSSM-like unification, and even correct for the $\sim3\%$ discrepancy found at two-loops, see e.g.~\cite{Alciati:2005ur}. We can express these effects in terms of corrections $\delta_i$ to the gauge coupling RGEs in the MSSM as follows 
\beq
\alpha_i^{-1}(m_Z) = 
&\alpha_{\rm GUT}^{-1}+
\frac{1}{2\pi}b_i^{\rm MSSM} {\rm log}\left(\frac{M_{\rm KK}}{m_Z}\right) + \delta_i~,
\label{00}
\eeq
where we define $\alpha^{-1}_{\rm GUT}$ to be the universal piece of the gauge kinetic functions $f_i$ of the SM gauge bosons. This is determined (up to a small contribution from the universal flux associated to the diagonal U(1) of U(5) \cite{Mayrhofer:2013ara}) by the local \kahler modulus $T$ which sets the volume of the GUT brane  
\beq
\alpha^{-1}_{\rm GUT}\simeq{\rm Re}\left[ T  \right]=2\mathcal{V}_{\rm GUT}~,
\eeq in units of $l_s\equiv 2\pi\sqrt{\alpha'}$, in the 10d Einstein frame.  
The quantity $\delta_i$ is the sum of all sources of corrections (specified shortly)
\beq
 \delta_i=\delta_i^{\rm MSSM} +  \delta_i^{\rm tree} + \delta_i^{\rm loop}~.
\eeq
Recall the $\beta$-function coefficients for the SM and MSSM are given by $b_i^{\rm SM}=\left(\frac{41}{10}, -\frac{19}{6},-7\right)$ and $b_i^{\rm MSSM}=\left(\frac{33}{5}, 1,-3\right)$. 
The corrections $\delta^{\rm MSSM}_i$ parameterise the departure of the SUSY scale $M_{\rm SUSY}$ from $m_Z$ 
\beq
\delta_i^{\rm MSSM}=\frac{1}{2\pi}\left(b_i^{\rm SM}-b_i^{\rm MSSM}\right){\rm log}\left(\frac{M_{\rm SUSY}}{m_Z}\right)~.
\label{susy}
\eeq
This can be trivially generalised to include splittings in the spectrum or variant (non-MSSM) spectra, as we comment on in Appendix \ref{S32}. 
Further, if the SUSY spectrum is perturbed from exact universality, to account for threshold corrections at the SUSY scale the sparticle scale $M_{\rm SUSY}$  in eq.~(\ref{susy}) should be replaced by \cite{Carena:1993ag}
\beq 
M_{\rm SUSY}~\longrightarrow~ m_{\widetilde{H}}\left(\frac{m_{\widetilde{W}}}{m_{\widetilde{g}}}\right)^{28/19}\left(\frac{m_{\widetilde{l}}}{m_{\widetilde{q}}}\right)^{3/19}\left(\frac{m_{H}}{m_{\widetilde{H}}}\right)^{3/19}\left(\frac{m_{\widetilde{W}}}{m_{\widetilde{H}}}\right)^{4/19}~,
\eeq
involving the various sparticle masses and the mass of the non-SM Higgs states $m_H$, assuming degenerate generations of squarks and sleptons. Thus such intermediate scale threshold effects are small for approximately universal spectra and henceforth we shall neglect such corrections.

Next we turn to the F-theory corrections, the tree-level piece is of the form \cite{Blumenhagen:2008aw,Mayrhofer:2013ara}
\beq
\delta_i^{\rm tree}=\frac{\gamma}{g_s} b^H_i~,
\label{bh}
\eeq
with $b^H_i=(3/5,1,0)$, equivalent to the $\beta$-function contribution due to $(1,2)_{-1/2}~+$ h.c. The factor $\frac{1}{g_s}$ should be identified with ${\rm Im}\left[\tau\right]$, in terms of the axio-dilaton $\tau\equiv is=i/g_s+C_0$ and $\gamma$ is a constant determined by the flux and vacuum configuration of the theory
\cite{Mayrhofer:2013ara}
\beq
\gamma=
\int_S\left[f_Y\wedge i^*B_--\frac{1}{2}f_Y\wedge f_Y-f_Y\wedge \widetilde{f}_S\right]~.
\label{ga}
\eeq
given in terms of the NS-NS form decomposed as $B_2\equiv B_+ + B_- = b^\alpha\omega_\alpha +b^a\omega_a$. Here $\widetilde{f}_S\equiv\left(f_S-\frac{2}{5}f_Y\right)$ in terms of the fluxes $f_Y$ and $f_S$ associated to, respectively, hypercharge and the diagonal U(1) factor of the U(5) GUT brane stack. The redefinition in terms of $\widetilde{f}_S$ is prudent as charge quantisation is, in one particular variant of the setting, realised by ensuring that $f_Y$ and $\widetilde{f}_S$ are integer quantised \cite{Mayrhofer:2013ara}. The total internal flux generator is given by
\beq
f=f_S\times{\bf 1}_5+f_Y
\left(\begin{array}{cc}
-\frac{2}{5} \times {\bf 1}_3 & 0\\
0 & \frac{3}{5} \times {\bf 1}_2
\end{array}\right)~.
\eeq
In order to avoid light exotics in the spectrum it is required \cite{Donagi:2008kj2} that $\int f_Y\wedge f_Y=-2$. For $f_Y\wedge i^*B_-=0$ the coefficient $\gamma$ is integer quantised, but otherwise it ranges over continuous values. We expect that typically $\gamma$ will be $\OO(1)$, in line with the fluxes taking natural values. For  the special case $i^*B_-= \widetilde{f}_S=0$, hence $\gamma=1$, this reduces to the setting studied by Blumenhagen \cite{Blumenhagen:2008aw}, with the gauge kinetic functions taking the (special) form
\beq
f_{i}&=T-\frac{s}{2}\int_S\left(f_S\wedge f_S+b^{H}_if_Y\wedge f_Y\right)~.
\label{fi}
\eeq

In another variant of this setting (see \cite{Mayrhofer:2013ara} for details), charge quantisation is realised when $f_Y$ and $f_S$ (rather than $\tilde{f}_S$) are integer quantised. It is then more convenient to give $\gamma$ as
\beq
\gamma=
\int_S\left[f_Y\wedge i^*B_--\frac{1}{10} f_Y\wedge f_Y-f_Y\wedge f_S\right]~.
\label{ga2}
\eeq
In fact, it is this variant which is arguably more relevant for F-theory GUTs. The reason is that the requirement of integer $f_S$ includes, in particular, the case of $f_S=0$. The latter avoids any potential issues with this diagonal U(1) in F-theory.\footnote{We thank Eran Palti for highlighting this latter variant and explaining its importance to us.}

The splitting of the gauge couplings from universality due to these tree level F-theory corrections is proportional to $b^{H}_i$. This can have crucial phenomenological consequences, which depend on the size of the effect and can be significant if ${\rm Im}\left[\tau\right]$ is large. More precisely, $1/g_s$ should be read as $e^{-\phi}$ averaged over the GUT brane. Simply making $g_s$ small is impossible in truly F-theoretic models due to the presence of an E$_6$ point (at which $\tau={\rm exp}(\frac{i \pi}{3})$, that is $g_s=\frac{2}{\sqrt{3}}$ \cite{Dasgupta:1996ij}), which is essential to obtain the top Yukawa coupling. However, regions of the GUT brane with large $e^{-\phi}$ are not excluded and thus large average values of $1/g_s$ are conceivable.

Finally, the loop-level F-theory corrections are of the form  \cite{Donagi:2008kj2} (see also Appendix \ref{apB})
\beq
\delta_i^{\rm loop}=
\frac{1}{2\pi}b_i^{5/6} {\rm log}\left(\frac{\Lambda}{M_{\rm KK}}\right)~,
\label{5/6}
\eeq
where $b_i^{5/6}=(5,3,2)$, analogous to the threshold correction coming from the off-diagonal parts of a {\bf 24}. Whilst there are also divergences associated to the massless modes running in loops cut-off at $\Lambda$ of the form $\frac{1}{2\pi}b_i^{\rm MSSM} {\rm log}\left(\frac{\Lambda}{M_{\rm KK}}\right)$ these are exactly cancelled by contributions from the KK excitations on the matter curves (this can be seen explicitly in the exposition of Appendix \ref{apB}) and thus eq.~(\ref{5/6}) gives the full correction.
Collecting these pieces, the total correction to the running  $\delta_i$, appearing in eq.~(\ref{00}), simplifies to
\beq
 \delta_i=\frac{1}{2\pi}\left(b_i^{\rm SM}-b_i^{\rm MSSM}\right) {\rm log}\left(\frac{M_{\rm SUSY}}{m_Z}\right)+\frac{\gamma}{g_s} b^H_i+\frac{1}{2\pi}b_i^{5/6} {\rm log}\left(\frac{\Lambda}{M_{\rm KK}}\right)~.
 \label{di}
\eeq

\subsection{Parameterising the precision of unification} 

The precision of unification can be parameterised, following \cite{Alciati:2005ur} (see also e.g.~\cite{Dolan:2011aq}), by fixing $\sin^2\theta_{\rm W}(m_Z)$ and $\alpha_{\rm EM}(m_Z)$ at the experimentally observed values and predicting the size of $\alpha^{-1}_3(m_Z)$ using the RGEs. Deviations from traditional TeV scale MSSM unification due to corrections $\delta_i$ can be parameterised in terms of a shift $\Delta_3$ in the predicted value of $\alpha^{-1}_3(m_Z)$
\beq
\alpha^{-1}_{3}(m_Z)=\alpha^{-1}_{3}\Big|_{\delta_i=0}+\Delta_3~.
\eeq
Here $\alpha^{-1}_{3}|_{\delta_i=0}$ is the predicted value of the QCD coupling at $m_Z$ in the absence of corrections (i.e.~setting $\delta_i=0$ in eq.~(\ref{00})). The corrections contribute to the shift $\Delta_3$ as follows \cite{Alciati:2005ur} 
\beq
\Delta_3    \equiv \frac{1}{7}\left(\Delta^{\rm SUSY}_3 +  \Delta^{\rm tree}_3 +  \Delta^{\rm loop}_3\right) =~ \frac{1}{7}\left(5 \delta_1 -  12\delta_2 +  7\delta_3\right)~.
\label{D3}
\eeq
For $\Delta_3=0$ TeV MSSM unification is preserved and, further, the aforementioned two-loop discrepancy can be reconciled with experiment if $\Delta_3\approx0.7$ \cite{Alciati:2005ur,Dolan:2011aq}. 
We can re-express eq.~(\ref{D3}) in terms of relative contributions as follows
\beq
\Delta_3^{\rm SUSY}  &  =
 \frac{19}{4\pi}{\rm log}\left(\frac{M_{\rm SUSY}}{m_Z}\right)~,\\
\Delta_3^{\rm tree}  &  
= -9\frac{\gamma}{g_s}~,
\\
\Delta_3^{\rm loop}  &  
=   \frac{3}{2\pi}{\rm log}\left(\frac{\Lambda}{M_{\rm KK}}\right)~.
\eeq

\begin{figure}[t!]
\begin{center}
\includegraphics[height=85mm]{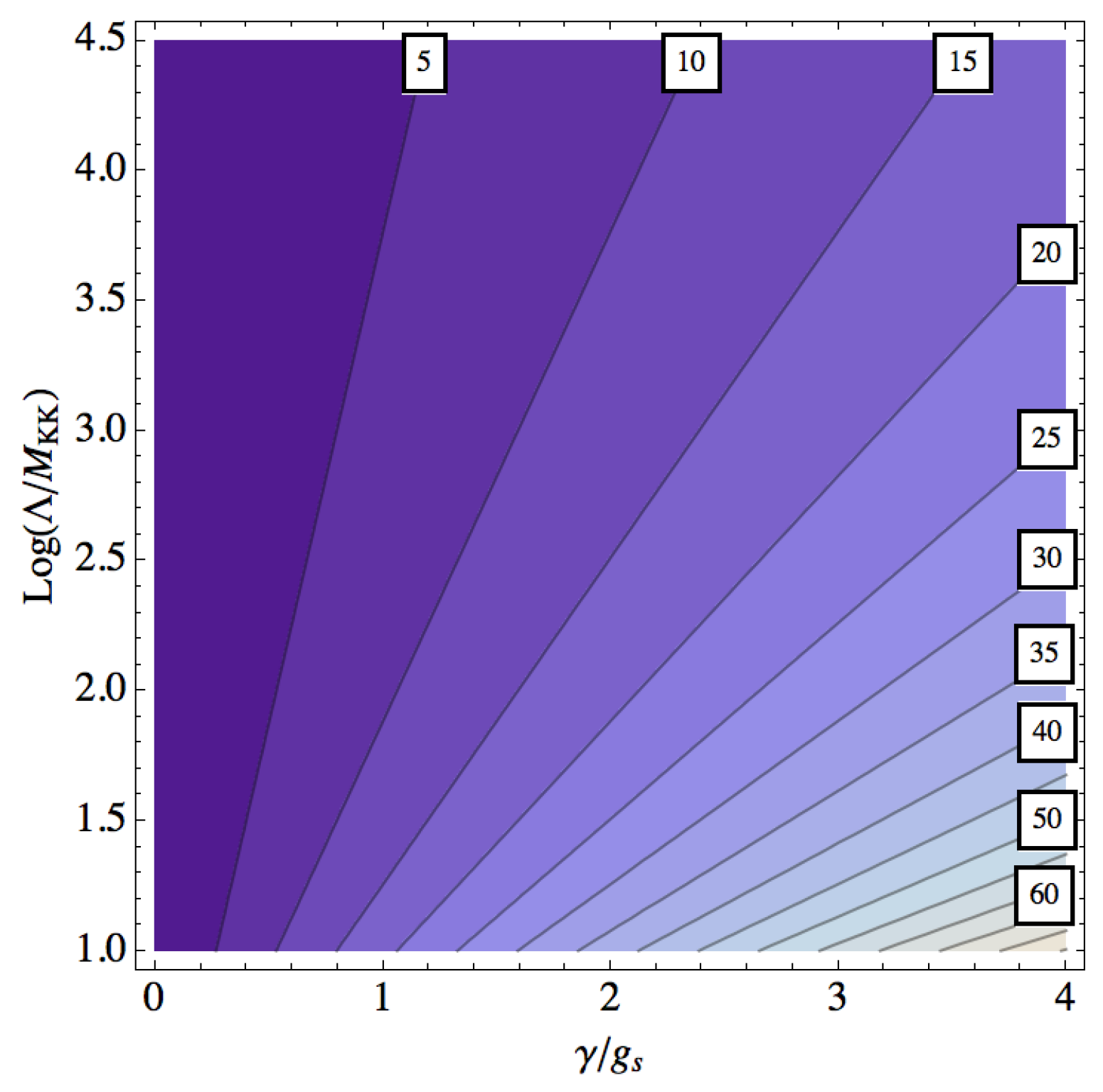}
\caption{
The magnitude of corrections $|\Delta_3^{\rm tree}/\Delta_3^{\rm loop}|$, as defined in eq.~(\ref{D3}), for varying prefactor $\gamma/g_s$ to $\delta_i^{\rm tree}$ and log factor ${\rm log}\left(\Lambda/M_{\rm KK}\right)$ of $\delta_i^{\rm loop}$. Observe that for the relevant parameter space the loop corrections are typically significantly less important than the tree level contributions.
 \label{tree-v-loop}}
\end{center}
\end{figure}

In the limiting case that the tadpoles are cancelled globally, with $\Lambda=T_s\mathcal{V}^{1/6} l_s$ in terms of the string tension $T=1/(2\pi\alpha')$, the logarithm dressing the UV correction is of order
 \beq
 {\rm log}\left(\frac{\Lambda}{M_{\rm KK}}\right)
 = {\rm log}\left(\frac{2\pi\mathcal{V}^{1/6}}{2\pi \mathcal{V}_{\rm GUT}^{-1/4}}\right)
= \frac{1}{6}~{\rm log}\left( \frac{\pi}{2\alpha_{\rm GUT}}\frac{M_{\rm Pl}^2}{M_{\rm GUT}^2}\right)
  \simeq4.4~,
\label{4.4}
 \eeq
 where we have used eq.~(\ref{qq}), (\ref{bb}) \& (\ref{bbb}) and, as we are working under the log, we have neglected unimportant factors of $g_s$.
  In the case of particular interest in which the tadpoles are cancelled locally, i.e.~$\Lambda\sim T_s \mathcal{V}_{\rm GUT}^{1/4}l_s$, the logarithmic enhancement is 
  \beq
 {\rm log}\left(\frac{\Lambda}{M_{\rm KK}}\right)=
 {\rm log} \left(\frac{2\pi\mathcal{V}_{\rm GUT}^{1/4}}{2\pi \mathcal{V}_{\rm GUT}^{-1/4}}\right)
 = \frac{1}{2}~{\rm log}\left(\frac{1}{2\alpha_{\rm GUT}}\right)
 \simeq1.2~.
\label{1.2}  \eeq
Thus the loop corrections are typically subdominant compared to the tree level corrections. Exponentiating, we note that local tadpole cancellation corresponds to $\Lambda/M_{\rm KK}\simeq3.3 $. We inspect the interplay of the two UV corrections in greater detail in Fig.~\ref{tree-v-loop}, which shows the relative magnitude due to the tree and loop level F-theory effects. In sizeable region of parameter space -- with local tadpole cancellation -- the loop level effects can be safely neglected. Further, this approximation also permits us to make contact with the existing literature, e.g.~\cite{Ibanez:2012zg,Blumenhagen:2008aw}.

Substituting the combined correction eq.~(\ref{di}) into the general form of $\Delta_3$ we obtain the deviation in the precision of unification in F-SU(5) High Scale SUSY, relative to the tree level SUSY GUT prediction
\beq
\Delta_3 &\simeq 0.7+ \frac{19}{28\pi}{\rm log}\left(\frac{M_{\rm SUSY}/m_Z}{7\times10^3}\right)
-\frac{9}{7}\left[\frac{\gamma/g_s}{1}-1\right]+\frac{3}{14\pi}{\rm log}\left(\frac{\Lambda/M_{\rm KK}}{3.3}\right)~.\\
\eeq
For the values indicated,  $\gamma/g_s\simeq1$ and $\Lambda/M_{\rm KK}\simeq3.3$ (chosen to give precision unification: $\Delta_3\simeq0.7$), the correction to $\alpha_3^{-1}(m_Z)$ coming from the F-theory one-loop piece is small $\Delta_3^{\rm loop}\simeq0.1$ compared to the tree $\Delta_3^{\rm tree}\simeq1.3$. 

 Incidentally, in the special case of (natural) TeV scale SUSY it is interesting to note that, for appropriate parameter choices, the F-theory corrections alone can in principle correct for the few percent deviation encountered at two loops in the MSSM. Such precision gauge coupling unification is found provided
\beq
\frac{\gamma}{g_s}-\frac{1}{6\pi}{\rm log}\left(\frac{\Lambda/M_{\rm KK}}{3.3}\right)
\simeq -0.1~.
\eeq
For instance, this relationship is satisfied for $\gamma\simeq-0.1$ with local tadpole cancellation or for $\gamma\simeq0$ with ${\rm log}(\Lambda/M_{\rm KK})\simeq3$. Note that negative values of $\gamma$, which corresponds to corrections that improve the precision of unification, are permitted by eq.~(\ref{ga}) \& (\ref{ga2}). The possibility of $\gamma<0$ has typically been neglected in the previous literature, which has mostly assumed the special choice of the flux embedding given in eq.~(\ref{fi}) for which $\gamma>0$. Thus, in addition to allowing for unification in High Scale SUSY, these F-theory corrections are another means of realising precision unification in the MSSM, distinct from existing mechanisms \cite{Carena:1993ag,Donkin:2010ta,Heckman:2011hu,Roszkowski:1995cn,Raby:2009sf,Hebecker:2004ce,Hall:2001xb,Hardy:2012ef}.

\subsection{The F-theory GUT scale in High Scale SUSY}
\label{S3.3}


We now come to the determination of $M_{\rm KK} \equiv M_{\rm GUT}$ from eq.~(\ref{00}). We assume that the $\delta_i$ are such that the correct low-energy values of the SM gauge couplings are reproduced. Furthermore, we work within the approximation that 1-loop supersymmetric unification without corrections works perfectly. This allows us to write
\beq
\alpha_i^{-1}(m_Z) = 
&\alpha_{\rm GUT}^{(0)\,\,-1}+
\frac{1}{2\pi}b_i^{\rm MSSM} {\rm log}\left(\frac{M^{(0)}_{\rm GUT}}{m_Z}\right)~,
\label{2}
\eeq
with uncorrected GUT scale $M^{(0)}_{\rm GUT}$ and GUT coupling $\alpha_{\rm GUT}^{(0)}$. Together with eq.~(\ref{00}) this gives 
\beq
\frac{1}{2\pi}b_i^{\rm MSSM} {\rm log}\left(\frac{M^{(0)}_{\rm GUT}}{M_{\rm GUT}}\right)+\left(\alpha_{\rm GUT}^{(0)\,\,-1}-\alpha_{\rm GUT}^{-1}\right)\quad=\quad
\delta_i\quad \equiv \quad\delta_i^{\rm MSSM} +  \delta_i^{\rm tree} + \delta_i^{\rm loop}~.
\eeq

If we form a linear combination of these three equations (labelled by $i=1,2,3$) with coefficients $\rho_i$, subject to the constraint $\rho_1+\rho_2+\rho_3=0$, then the term involving GUT couplings drops out and 
$M_{\rm GUT}$ can be read off. Since the overall normalisation of the $\rho_i$ is irrelevant, there is one meaningful degree of freedom in this choice. In the spirit of Blumenhagen \cite{Blumenhagen:2008aw} we use this freedom to ensure that $\delta_i^{\rm tree}$ also drops out, taking e.g. $\rho_i=\{1,-3/5,-2/5\}$. This gives  \beq
\frac{1}{2\pi}B^{\rm MSSM} {\rm log}\left(\frac{M^{(0)}_{\rm GUT}}{M_{\rm GUT}}\right)=\left(\delta_1^{\rm MSSM}+\delta_1^{\rm loop}\right)
-\frac{3}{5}\left(\delta_2^{\rm MSSM}+\delta_2^{\rm loop}\right)
-\frac{2}{5}\left(\delta_3^{\rm MSSM}+\delta_3^{\rm loop}\right)\,,
\label{1}
\eeq
with
\beq
B^{\rm MSSM}\equiv b_1^{\rm MSSM}-\frac{3}{5}b_2^{\rm MSSM}-\frac{2}{5}
b_3^{\rm MSSM}=\frac{36}{5}~.
\label{B}
\eeq

Defining the analogous SM quantity $B^{\rm SM}=44/5$ and using eqs.~(\ref{susy}) and (\ref{5/6}), we can give the corrected GUT scale in a slightly more compact form, or (in the last line) completely explicitly: 
\beq
{\rm log}\left(\frac{M_{\rm GUT}}{M^{(0)}_{\rm GUT}}\right)
&=-\frac{1}{B^{\rm MSSM}}\Big[\left(B^{\rm MSSM}-B^{\rm SM}\right)
{\rm log}\left(\frac{M_{\rm SUSY}}{m_Z}\right)\\
& \hspace{45mm} +\left(b_1^{5/6}-\frac{3}{5}b_2^{5/6}-\frac{2}{5}b_3^{5/6}\right){\rm log}\left(\frac{\Lambda}{M_{\rm KK}}\right)\Big]\\
&=-\frac{2}{9}~
{\rm log}\left(\frac{M_{\rm SUSY}}{m_Z}\right)
-\frac{1}{3}~{\rm log}\left(\frac{\Lambda}{M_{\rm KK}}\right)~.
\label{373'}
\eeq 
Note that here and below $\Lambda/M_{\rm KK}$ is given by either eq.~(\ref{4.4}) or eq.~(\ref{1.2}), independently of $M_{\rm GUT}$ on the LHS, which is determined by these equations. 

As already emphasized in \cite{Ibanez:2012zg}, raising the masses of the superpartners leads to a reduction in the GUT scale. We see from eq.~(\ref{373'}) that the loop correction further enhances this effect. 

Exponentiating, we obtain\footnote{
Analogous 
expressions appear in the literature for standard GUTs with the Split SUSY spectrum \cite{Giudice:2004tc}. In Appendix \ref{S32} we provide a general expression for two intermediate mass scales and with the $\beta$-function coefficients unspecified.}
\beq
M_{\rm GUT}=
M_{\rm GUT}^{\rm (0)}\times
\left(\frac{m_{Z}}{M_{\rm SUSY}}\right)^{2/9}
\left(\frac{M_{\rm KK}}{\Lambda}\right)^{1/3}~.
\label{111}
\eeq
\begin{figure}[t!]
\begin{center}
\includegraphics[height=70mm]{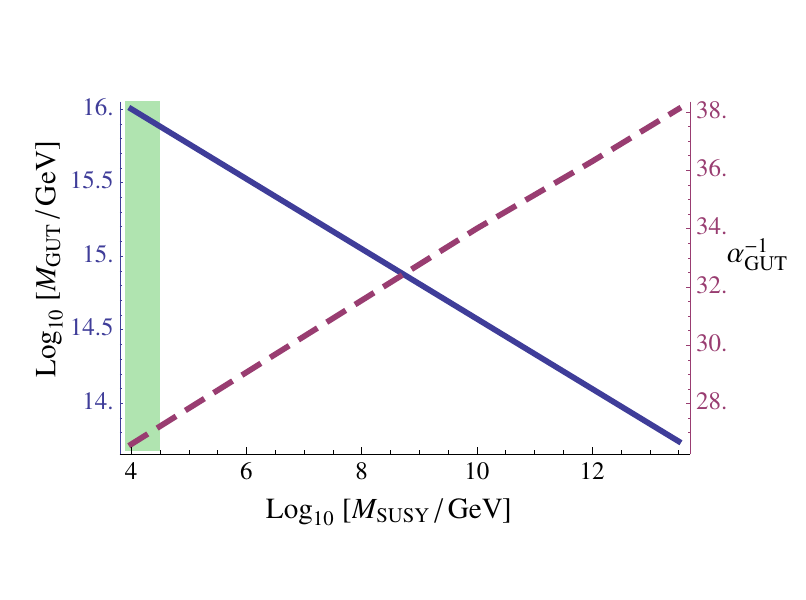}
\caption{
The GUT scale (blue, solid) and GUT coupling (red, dashed) determined by running the gauge coupling to the point of F-theory unification, where we assume local tadpole cancelation as given in eq.~(\ref{1.2}). The shaded region satisfies the proton decay bound $\tau(p\rightarrow\pi^0e^+)>2.5\times10^{41}$s, assuming no suppression in the proton decay rate, which we discuss further in subsequent sections.
 \label{Fig2}}
\end{center}
\end{figure}
Minimizing the loop effect by assuming local tadpole cancellation, as given in eq.~(\ref{1.2}), the GUT scale is numerically given by
\beq
M_{\rm GUT} \simeq 4.25\times10^{15}~{\rm GeV}
\left(\frac{10^{5}~{\rm GeV}}{M_{\rm SUSY}}\right)^{2/9}
\left(\frac{3.3}{\Lambda/M_{\rm KK}}\right)^{1/3}
~.
\label{aa}
\eeq

For later use in the proton decay sections, we also give the value of $\alpha_{\rm GUT}$. This value is somewhat uncertain since we neglected GUT-universal high scale corrections. We also do not want to go into the details of which specific combinations of $\alpha_1$, $\alpha_2$ and $\alpha_3$ at the high scale are relevant for the couplings of the $X,Y$ gauge bosons. Instead, we simply use the high scale value of $\alpha_3$, as determined from 
eq.~(\ref{00}):
\beq
\alpha_{\rm GUT}^{-1}
&\simeq\alpha_3^{-1}(m_Z)+
\frac{1}{2\pi}\left[
b_3^{\rm MSSM}\log\left(\frac{M_{\rm SUSY}}{M_{\rm GUT}}\right)
+b_3^{\rm SM}\log\left(\frac{m_{Z}}{M_{\rm SUSY}}\right)
\right]\\
&
\simeq 28+\frac{10}{6\pi}\log\left(\frac{M_{\rm SUSY}}{10^5~{\rm GeV}}\right)~,
\label{gg}
\eeq
where we have neglected the subleading UV corrections $\delta^{\rm tree}_i$ and $\delta^{\rm loop}_i$.

In Fig.~\ref{Fig2} we show $M_{\rm GUT}$ and $\alpha_{\rm GUT}^{-1}$ as a function of $M_{\rm SUSY}$ determined by running the gauge couplings from the weak scale via eq.~(\ref{aa}) \& (\ref{gg}). The shaded region indicates roughly the parameter space in which proton decay bounds are satisfied, assuming no additional suppressions in the proton decay rate, which we discuss in detail in Sect.~\ref{S3b} \& \ref{S4}.


\section{Dimension-6 proton decay in F-theory GUTs}
\label{S3b}

We focus on the most robust and, in particular, SUSY-breaking-independent 
proton decay predictions: those due to dimension-6 operators induced by $X,Y$
gauge bosons.  One of the suggestions of \cite{Ibanez:2012zg} was that the relevant couplings of the $X,Y$ bosons to SM matter can easily be exponentially suppressed in F-theory GUTs. For this, the $X,Y$ bosons have to be localised 
(due to gauge flux) away from the matter curves, such that their wavefunction overlap with matter fields is negligible. Similar proposals have been made in the context of field-theoretic higher-dimensional GUT models \cite{Hamada:2012wj,Kakizaki:2013ila}. However, these arguments neglect the effects of $X,Y$ modes at higher Landau levels, which also mediate proton decay and are less localized than the lowest states. Here we argue that, unless the construction is very special, such higher modes will always induce proton decay without exponential suppression. We estimate the corresponding rates.\footnote{Our analysis is not restricted to high-scale SUSY and equally applies to F-theory GUTs with TeV-scale SUSY.}


\subsection{Localisation on a compact surface}
\label{SS2}

To set the stage, we first recall a few basic formulae concerning wavefunctions in the presence of flux \cite{Bachas:1995ik}. While we are actually interested in an 8d gauge theory, our main qualitative point can be made in 6d. Thus, we simplify our setting by starting from an 8d theory compactified on $T^4$, but immediately let two of the radii be so small that the system reduces to a 6d gauge theory on $T^2$. We use 
co-ordinates $(x_5,x_6)$, such that $x_i=x_i+2\pi R_i$ for $i=5,6$ and we define $L_i=2\pi R_i$. 

Following \cite{Bachas:1995ik}, in the presence of a constant background field $F_{56}\equiv\partial_5 A_6-\partial_6 A_5=H$ the component fields can be parameterised as 
\beq
A_5=0~,  \hspace{20mm} A_6=H x_5~.
\eeq 
The field $A_\mu$ is periodic in $x_5$ up to a gauge transformation, 
\beq 
A_\mu (x_5+L_5)=A_\mu(x_5)-ie^{-i\theta}\partial_\mu e^{i\theta}~,
\eeq
with $\theta=HL_5x_6$. Demanding that this be single-valued in $x_6$-direction leads to the Dirac quantisation condition
\beq
H=\frac{2\pi N}{L_5 L_6},
\label{H}
\eeq
where $N\equiv\frac{1}{2\pi}\int_{T^2}F_{56}$ is an integer which counts the flux number.
The 4d spectrum of the model is determined by the Laplacian, 
\beq
(D_{5}^2+D_6^2)\Phi_n=
\left[\partial_5^2+(\partial_6+H x_5)^2 \right]\Phi_n=M_n^2\Phi_n~.
\label{lap}
\eeq

For example, the set of ground state wavefunctions $\Phi_{0}^{(j)}$ can be expressed as \cite{Bachas:1995ik}
\beq
\Phi_0^{(j)}&= \sum_{m=-\infty}^\infty
\exp\left[-\frac{\pi N}{L_5L_6}\left(x_5-\left(m +\frac{j}{N}\right)L_5\right)^2\right]~\exp\left[2\pi i\left(N m +j\right)\frac{x_6}{L_6}\right]~.
 \label{e1}
\eeq
Note that the fields respect the periodicity conditions
\beq
\Phi(x_5+L_5,x_6) =e^{i\theta}\Phi(x_5,x_6)~,
\hspace{15mm}
\Phi(x_5,x_6+L_6) =\Phi(x_5,x_6)~.
\label{per}
\eeq
By relabelling $m$ it can be seen that $\Phi^{(j)}=\Phi^{(j+N)}$ and thus that the spectrum features an $N$-fold degeneracy. 
The ground state wavefunctions are approximate Gaussians (with respect to the co-ordinate $x_5$), distributed evenly within the interval $[0,L_5]$ and with width
\beq
d=\sqrt{\frac{L_5L_6}{2\pi N}}~.
\label{d}
\eeq
Now, identifying these wavefunctions with $X,Y$ gauge bosons and asking for an exponential suppression of proton decay, our aim must be to find regions of the compact space where all these (ground state) wavefunctions vanish. Such regions obviously emerge if 
\beq
\frac{Nd}{L_5}=\sqrt{\frac{L_6}{L_5}\cdot \frac{N}{2\pi}}\ll 1\,.
\eeq
We see that we are forced into the effectively one-dimensional regime 
$L_6/L_5\ll 1$ (a `thin' torus). This discussion is similar to that of \cite{Hayashi:2009bt} in the context of SM Yukawa couplings.

Most importantly, we have learned that compactness of the space and degeneracy of the Landau levels are essential. They reveal constraints which are relevant for proton decay suppression but are not visible in a `local' analysis, i.e. working on $\mathbb{C}$ (or $\mathbb{C}^2$).


\subsection{Modification of proton decay rates}
\label{SS3}

We now set up some notation for writing down the proton decay rate, including the effects of both ground state and higher Landau levels of $X,Y$ bosons. The simplifying assumptions of a $T^4$ GUT brane or of a 6d effective theory are not needed for the moment. 
We start from the 8d action
\beq
\mathcal{S}=\int_{\mathbb{R}^4\times S} \overline{\Psi}i\slashed{D}\Psi+\frac{1}{4g_8^2}A_M\partial^2A^M+\cdots~,
\label{e.a}
\eeq
where $\Psi$ is an 8d spinor and $M=\{\mu,m\}$. Here $\mu=0,1,2,3$ are 4d Lorentz indices and $m=5,6,7,8$ label the internal dimensions. The 8d spinor is decomposed, according to $\Psi=\psi(x^\mu)\varphi(y^m)$, into a 4d Lorentz spinor $\psi$ and the internal spinor $\varphi$, which is a zero mode of the internal Dirac operator (cf. e.g.~\cite{Camara:2011nj,Hayashi:2009bt,Palti:2012aa,Camara:2013fta}): 
\beq
\slashed{D}_{\rm int.}\varphi=0\,,\qquad\mbox{with}\qquad \slashed{D}=\slashed{D}_4 + \slashed{D}_{\rm int.}\,.
\eeq
The field $A_M$ can be decomposed as $\{A_M\}=\{A_\mu=A^n_\mu(x)\Phi_n(y),\cdots\}$,  where $\Phi_n(y)$ are modes of an internal scalar. 

Starting with the 8d fermion states, the first term of eq.~(\ref{e.a}) decomposes to give
\beq
\int_{\mathbb{R}^4\times S} \overline{\Psi}i\slashed{D}\Psi=\int_S\varphi^\dagger\varphi\int_{\mathbb{R}^4} \overline{\psi}i\slashed{\partial}_4\psi
+\sum_{n=0}^\infty~\int_S\varphi^\dagger\varphi\, \Phi_n\int_{\mathbb{R}^4} \overline{\psi}i(iA^n_\mu\gamma^\mu)\psi~,
\label{act1}
\eeq
where the sum runs over the KK modes.
The prefactors can be identified as the normalisation of the kinetic function and the coupling to gauge bosons, respectively. Next, for the 8d gauge bosons we have
\beq
\int_{\mathbb{R}^4\times S} \frac{1}{4g_8^2}A^n_M\partial_{8D}^2A_n^M=\frac{1}{4g_8^2}\int_{S} \Phi_n^2
\int_{\mathbb{R}^4} A^n_\mu(\partial_{4D}^2+M_n^2)A_n^\mu+\cdots\,,\label{act2}
\eeq
where $M_n^2$ is the eigenvalue of the field $\Phi_n(y)$ under the internal Laplace operator. For the SM subgroup, it runs over zero and the higher KK modes while, for the $X,Y$ gauge bosons, it labels the Landau levels.

Now, just from the parts of the action displayed in eqs.~(\ref{act1}) and 
(\ref{act2}), it is easy to estimate the coefficient of the 4-fermion operator $\sim(\overline{\psi}\psi)^2$ which is induced by integrating out the $n$th mode of the vector field with mass $M_n$. This `toy model' coefficient reads 
\beq
F_n^{\rm toy}=\frac{4g_8^2 \int_S \varphi^\dagger\varphi \Phi_n}{(\int_S |\varphi|^2)^2 (\int_S \Phi_n^2)M_n^2 }~.\label{ftoy}
\eeq
It is built from four fermion-normalization-factors, two vertex-normalization-factors (cf.~eq.~(\ref{act1})) and one zero-momentum vector propagator (cf.~eq.~(\ref{act2})) in a self-explanatory manner. We do not worry about the overall numerical prefactor, taking eq.~(\ref{ftoy}) as the definition of the normalization of $F_n^{\rm toy}$.

The generalization required for actual proton decay is minimal: We now interpret $\Psi$ as a field in the ${\bf 10}$ of $SU(5)$. The relevant internal wavefunctions, in SM language, then read $\varphi=\{\varphi_Q,\varphi_u,\varphi_e\}$. The familiar  ${\bf 10}$-${\bf 10}$ proton decay operator arises from a diagram where the $X,Y$ bosons couple to $(Q,u)$ at one side and to $(Q,e)$ at the other. This leads to an operator coefficient 
\beq
F_n=
\frac{4g_8^2(\int_S \varphi^\dagger_Q\varphi_u \Phi_n)^\dagger(\int_S \varphi^\dagger_Q\varphi_e \Phi_n)}{(\int_S |\varphi_Q|^2)(\int_S |\varphi_u|^2)^{1/2}(\int_S |\varphi_e|^2)^{1/2}
(\int_S \Phi_n^2)M_n^2}~.\label{fn}
\eeq
We clearly see, in the numerator, the relevant triple overlaps of the wavefunctions $\Phi_n$ and $\varphi_{Q,u,e}$ of the states involved in the proton decay operator. Further, the 8d gauge coupling can be re-expressed in terms of the 4d coupling as follows $\frac{1}{g_4^2}=\frac{1}{g_8^2}(\int_S 1)$. If we assume that $\varphi_Q= \varphi_u= \varphi_e$ and $\Phi$ are constant, the result must reduce to the 4d case provided we identify $M_n$ with $M_{\rm GUT}= M_{X,Y}$. Explicitly, with these assumptions, eq.~(\ref{fn}) gives 
\beq
F_{\rm 4D} = \frac{4g_4^2}{M_{\rm GUT}^2}~,
\eeq
which we take as the definition of the corresponding 4d-GUT operator coefficient. 

It is convenient to consider the ratio of the contribution of the $n$th Landau level mode and the conventional contribution of 4d GUT $X,Y$ bosons
to proton decay. In this ratio, the arbitrary normalization of $F_n$ drops out. Thus, we have the unambiguous expression 
\beq
\beta_n \equiv\frac{F_n}{F_{\rm 4D}}=
\frac{(\int_S \varphi^\dagger_Q\varphi_u \Phi_n)^\dagger(\int_S \varphi^\dagger_Q\varphi_e \Phi_n)(\int_S 1)}
{(\int_S |\varphi_Q|^2)(\int_S |\varphi_u|^2)^{1/2}(\int_S |\varphi_e|^2)^{1/2}(\int_S \Phi_n^2)}\cdot\frac{M_{\rm GUT}^2}{M_n^2}~.
\eeq
Note that $\beta_n$ has no explicit dependence on the normalisation of the fields or the metric on $S$ since both fields and integrals appear an equal number of times in numerator and denominator. The F-theory proton decay rate 
is now obtained by summing over all $\beta_n$, modifying the 4d decay rate accordingly:
\beq
\Gamma= \Gamma_{\rm 4D}\left(\sum_{n=0}^\infty \beta_n\right)^2=
\frac{\mathcal{C} m_p^5
\alpha_{\rm GUT}^2}{M_{\rm GUT}^4} \left(\sum_{n=0}^\infty \beta_n\right)^2~.
\label{e.b}
\eeq
The quantity $\mathcal{C}\approx2.16$ accounts for hadronic physics and factors out in our analysis. We will discuss this factor in greater detail in Sect.~\ref{S4}. We finally note that the Landau level spectrum, 
\beq 
\left(\frac{M_n}{M_{\rm GUT}}\right)^2\equiv\left(\frac{M_n}{M_0}\right)^2= 2\left(n+\frac{1}{2}\right)~,
\label{31}
\eeq
which is an important ingredient in $\beta_n$, differs significantly from the 
more familiar KK mass spectrum, $M_n^{\rm KK}/M_1^{\rm KK}=n$. In eq.~(\ref{31}) we have identified the GUT scale with the mass of the lowest Laundau level.  This overrides the identification with the lowest KK mass used in earlier sections, where we had not yet discussed dimensional reduction with flux. The difference is a numerical factor of little relevance in the leading-log unification analysis performed above.

We now simplify to the case where the GUT brane factorizes in two 2D spaces,
$S=A\times B$, and both the $\varphi$ and $\Phi$ modes are constant along $B$. For example, the GUT brane could be $T^4=T_1^2\times T_2^2$, with the second two-torus being much smaller, leading to the effectively 6-dimensional setting described earlier. We now have 
\beq
\beta_n=\frac{(\int_A \varphi^\dagger_Q\varphi_u \Phi_n)^\dagger(\int_A \varphi^\dagger_Q\varphi_e \Phi_n)(\int_A 1)}
{(\int_A |\varphi_Q|^2)(\int_A |\varphi_u|^2)^{1/2}(\int_A |\varphi_e|^2)^{1/2}(\int_A \Phi_n^2)}\cdot\frac{M_{\rm GUT}^2}{M_n^2}~.
\eeq
Further, for $\varphi_Q\simeq\varphi_u\simeq\varphi_e$ and if $|\varphi_Q|^2$ is strongly localised at the point $x_0\in A$ we obtain
\beq
\beta_n =
\frac{|\Phi_n(x_0)|^2 \mathcal{V}_A}{\int_A |\Phi_n|^2}\cdot\frac{1}{2n+1}~,
\label{eq1}
\eeq
where $\mathcal{V}_A\equiv \int_A 1$ is the volume of $A$. This is the case which we shall study henceforth. One can think of this as a class of models 
in which the matter curve is identified with $B$ (hence being localized at a point in $A$) and where the flux is experienced only on $A$.
 Note that the first factor in eq.~(\ref{eq1}) characterizes the relative strength of the $n$th Landau level mode at the point $x_0$, and the second factor provides the appropriate mass suppression.


\subsection{Proton decay via higher modes}
\label{SS4}


Next we utilize the general expression for the proton decay rate,  eq.~(\ref{eq1}), to argue that even in the case that the ground state wavefunction overlap with the SM fields is negligible, some higher modes 
will have substantial overlap with the matter curve. Consequently, proton decay is not exponentially suppressed. 

\begin{figure}[t!]
\begin{center}
\includegraphics[height=45mm]{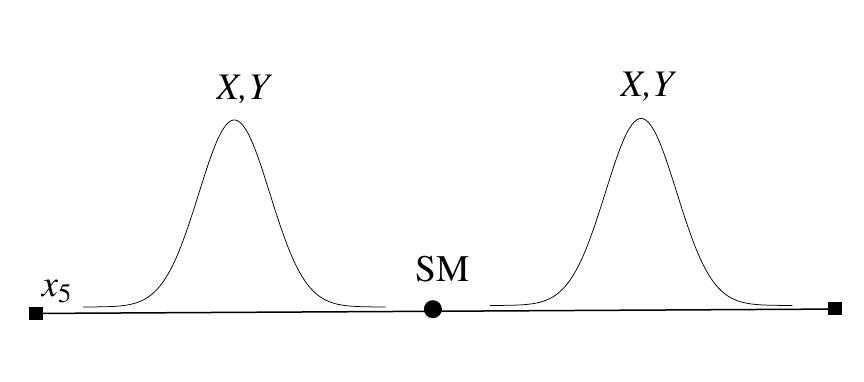}
\caption{\label{fig1} 
For flux $N=2$ the $X,Y$ eigenvalues have a two-fold degeneracy. The $x_5$-dimension is compactified to $S^1$ by identifying the end points (indicated by the squares). For the case that the SM matter curve (marked `SM') is equidistant to the two peaks, the proton decay mediated via the ground state is maximally suppressed.
}
\end{center}
\end{figure}

We take $A$ to be a rectangular 2-torus with volume $L_5\times L_6$. As we already know, the most promising geometry is the effectively 1-dimensional limit where $L_6\ll L_5$. Thus, we now focus on a 5d theory compactified on $S^1$ with volume $L_5$ and, for simplicity, define $x_5\equiv x$. The $N$ ground state wavefunctions of $X,Y$ bosons are approximate Gaussians, cf. eq.~(\ref{e1}), which are localized at $N$ equally spaced positions along $x$. We fix the origin to coincide with one 
of these peaks. More specifically, we choose the peak which is closest to the matter curve. This allows us to consider only the patch $|x|\leq \frac{L_5}{2N}$
in what follows.
The $X,Y$ boson ground state wavefunction near the origin is described by
\beq
\Phi_{0}\sim\exp\left(-\frac{x^2}{2d^2}\right)~.\label{phi0}
\eeq 
The next peak is at $x=\frac{L_5}{N}$. It is clear that maximal suppression of proton decay due to exchange of the $X,Y$ ground state is achieved when the 
matter curve (which now corresponds to a point) is at exactly half-way between adjacent peaks, i.e.~at $x=\frac{L_5}{2N}$. This is illustrated in Fig.~\ref{fig1}.

We are interested in the effects of the higher Landau levels. Whilst the corresponding wavefunctions could be obtained by solving eq.~(\ref{e1}) in the limit $L_6\ll L_5$, there is a much simpler approach, based on the analogy with the quantum harmonic oscillator. We start from the 1d internal Klein-Gordon-type equation derived by focusing on the behaviour of the field in the $x_5$-direction
\beq
\left[-\partial_x^2-M(x)^2\right]\Phi_n(x)=m_n^2\Phi_n(x)~,
\eeq
where $M(x)=Hx$, with $H$ defined in eq.~(\ref{H}). Here we are already working in the thin torus limit $L_6\ll L_5$ and focusing on the local patch $|x|\le \frac{L_5}{2N}$. This can be re-expressed as
\beq
\left[(-\partial_x+Hx)(\partial_x+Hx)+H\right]\Phi_n(x)=m_n^2\Phi_n(x)\,.
\label{lap2}
\eeq
The function $\Phi_0$ is a solution of $(\partial_x+Hx)\Phi_0=0$, hence $\Phi_0\sim\exp\left(-H x^2/2\right)$. This is consistent with eq.~(\ref{phi0}) since $H=1/d^2$. By analogy to the harmonic oscillator spectrum, 
\beq
m_n^2=2H\left(n+\frac{1}{2}\right)~.
\label{ll}
\eeq

To find the full set of eigenfunctions of this system we map eq.~(\ref{lap2}) to the harmonic oscillator (see e.g.~\cite{Bachas:1995ik}), which suggests the following ladder operators
\beq
a=\sqrt{\frac{d}{2}}\left(x_5+d^2\frac{{\rm d}}{{\rm d}x_5}\right)~,
\hspace{5mm}
a^\dagger=\sqrt{\frac{d}{2}}\left(x_5-d^2\frac{{\rm d}}{{\rm d}x_5}\right)~.
\eeq
The higher modes are then obtained by acting on the ground state wavefunction $\Phi_{0}$, which we normalise as follows
\beq
\Phi_{0}=\frac{1}{\pi^{1/4}\sqrt{d}} \exp\left(-\frac{x^2}{2d^2}\right)~.\label{phi0'}
\eeq 
With the ladder operators we obtain
\beq
\Phi_n(x)= \frac{1}{\pi^{1/4}\sqrt{2^n n!d}}~H_n\left(\frac{x}{d}\right)\exp\left(-\frac{x^2}{2d^2}\right)~,
\label{psin}
\eeq
in terms of the Hermite polynomials $H_n$. Each mode $\Phi_n$ satisfies the normalisation condition
\beq
\int^\infty_{-\infty}|\Phi_n(x)|^2~{\rm d}x=1~. 
\label{norm}
\eeq
As higher Landau level modes experience more acutely the structure of the internal space, our approximation breaks down at large $n$ in curved geometries. However, high $n$ corresponds to the UV regime of the theory and flux, being an IR effect, should not affect this UV sector. Thus, we still expect that UV degrees of freedom of $X,Y$ bosons contribute to proton decay, even if our Landau level estimate breaks down. Nevertheless, this is just an expectation and the combined effects of flux and curvature clearly requires further study.

An explicit form for the suppression factor can be obtained by substituting eq.~(\ref{psin}) into eq.~(\ref{eq1}), leading to
\beq
\beta_n =
\frac{\mathcal{V}_A}{\int_A |\Phi_n|^2}\cdot\frac{1}{2n+1}\cdot \frac{1}{\pi^{1/2}2^n n!d}\left[H_n\left(\frac{x_0}{d}\right)\right]^2\exp\left(-\frac{x_0^2}{d^2}\right)~.
\label{eq3}
\eeq
We can now consider the relative enhancement of a given excited mode to the ground state  \beq
\frac{\beta_n}{\beta_0} =
\frac{\int_A |\Phi_0|^2}{\int_A |\Phi_n|^2}\cdot\frac{1}{2n+1}\cdot \frac{1}{2^n n!}\left[H_n\left(\frac{x_0}{d}\right)\right]^2~.
\label{eq4}
\eeq
As this expression has no explicit volume dependance, we can work in the limit $L_5\rightarrow\infty$, in which case
\beq
\int_A |\Phi_n|^2\approx\int^{\infty}_{-\infty} ~|\Phi_n(x)|^2~{\rm d}x
=1~,
\eeq
and the normalisation factors drop out. Thus the relative enhancement is given by
\beq
\frac{\beta_n}{\beta_0} =
\frac{1}{2n+1}\cdot \frac{1}{2^n n!}\left[H_n\left(\frac{x_0}{d}\right)\right]^2~.
\label{eq5}
\eeq

\begin{figure}[t!]
\begin{center}
\includegraphics[height=45mm]{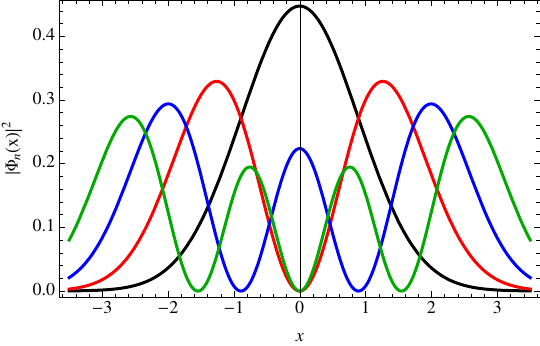}
\hspace{4mm}
\includegraphics[height=45mm]{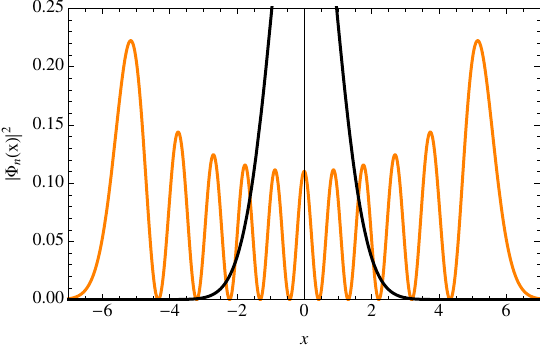}
\caption{ 
Ground state $\Phi_0$  (black) for $N,L_6=1$ and $L_5=10$, accompanied by $\Phi_1$ (red), $\Phi_2$ (blue), and $\Phi_3$  (green) in the left panel and $\Phi_{10}$ (orange) in the right panel.
\label{left}}
\end{center}
\end{figure}

The Landau levels with significant overlaps with the SM matter curves are those with maxima near $x_0$. The position of the maximum $x_*^{(n)}$ of each Landau level (approximately given by the outmost peak of the associated Hermite polynomial) can be found by solving $\Phi_n^\prime=0$. For the first few modes the maxima are located at
 \beq
 x_*^{(1)}=\pm d~,
 \hspace{5mm}
 x_*^{(2)}=\pm \sqrt{\frac{5}{2}}~d~,
  \hspace{5mm}
 x_*^{(3)}\approx\pm 2d~,
\label{pe}
 \eeq
 as can be seen from inspection of Fig.~\ref{left}.
The corresponding peak values can be calculated by evaluating $\Phi_n^2\left(\frac{x_*}{d}\right)$.
 In the scenario that proton decay is dominantly mediated by  higher (large $n$) modes, the corresponding wavefunctions are similar in form to cosine functions. This can be seen \cite{math2} by looking at the asymptotic form  (for $|x|<\sqrt{2n}$) 
 \beq
 H_n (x)e^{-x^2/2} \sim \left[\left(2ne^{-1}\right)^{n/2}\sqrt{2}\left(1-\frac{x^2}{2n}\right)^{-1/4}\right]\cos\left(x\sqrt{2n}-\frac{n\pi}{2}\right)~.
 \label{asym}
 \eeq
The function rises sharply at $\sqrt{2n}$, this is the position of the outermost peak and the global maximum (the unbounded behaviour is simply an artifact of the approximation).  Outside of the domain $|x|<\sqrt{2n}$ the function is heavily suppressed.
For large $n$ the outermost peak (the global maximum) occurs  \cite{math} at 
\beq
x_*^{(n)}\sim d\sqrt{2n}~. 
\eeq

The mode $\Phi_{\hat{n}}$ which dominantly contributes to the proton decay rate corresponds to the wavefunction which is peaked in the vicinity of the matter curve $x_0$ and satisfies the condition $x_*^{(\hat{n})}\sim x_0$. Therefore if the ground state is maximally separated from SM matter, $x_0=\frac{L_5}{2N}$, then 
 \beq
\hat n\sim \frac{L_5^2}{8N^2d^2} \sim\frac{\pi}{4N}\frac{L_5}{L_6}~.
\label{hh}
\eeq
Clearly $\hat n$ must take integer values, so the RHS should be appropriately rounded. 
 By plotting the first few modes, see Fig.~\ref{left}, it is not difficult to convince oneself that the peaks of the higher modes are dense over the GUT surface. Thus, although the wavefunction for individual modes (e.g.~the ground state) can be negligible at the position of the SM fields, the full proton decay rate typically does not receive a significant suppression.

 To illustrate this point we shall give a concrete example in which the ground state wavefunction overlap is exponentially suppressed but the third Landau level $\Phi_3$ has a significant wavefunction overlap. Suppose, as before, that the matter curve is equidistant between two peaks of the wavefunction  $x_0=\frac{L_5}{2N}$.
 The exponential suppression which arises for the ground state can be seen by evaluating $\Phi_0$ at the position of the matter curve $x_0$. The modification factor  eq.~(\ref{eq1})  is given by 
\beq
\frac{\beta_0 (x_0)}{ \mathcal{V}_A}
=\Phi_0^2\left(x_0\right)
\sim\exp\left[-\left(\frac{\pi L_5}{2NL_6}\right)\right]~.
\label{eq2}
 \eeq
  Note that this exponential factor can  easily be small and it appears squared in the proton decay rate $\Gamma$. This inverse exponential factor is present in all of the $\beta_n$, however it is negated if $\frac{x_0}{|x_*|}\approx1$, where $x_*$ is the position of the wavefunction maximum.  Now consider the special case where the matter curves are located at $L_5=8NL_6/\pi$, and hence $x_0=2d$. In this scenario the decay rate due to the ground state receives an exponential suppression. However, as noted in eq.~(\ref{pe}), the maximum of $|\Phi_3(x)|^2$ is found at $x_*\approx2d$, and thus the factor $\beta_3$ does not exhibit this large suppression. Consequently, by comparison to eq.~(\ref{eq5}), proton decay mediated by the third Landau level is greatly enhanced over the ground state channel 
  \beq
\Gamma_3=\left(\frac{\beta_3(2d)}{\beta_0(2d)}\right)^2\Gamma_0=\left(\frac{1}{2\times3+1}\right)^2 \left(\frac{100}{3}\right)^2\Gamma_0
~,
  \eeq
where $\Gamma_n$ is the partial decay rate due to the $n$th Landau level. This special case demonstrates that whilst the decay rate of the ground state may be exponentially suppressed, the higher modes which overlap the matter curve will be significant.

\subsection{Numerical analysis of the proton decay rate}

\begin{figure}[t!]
\begin{center}
\includegraphics[height=45mm]{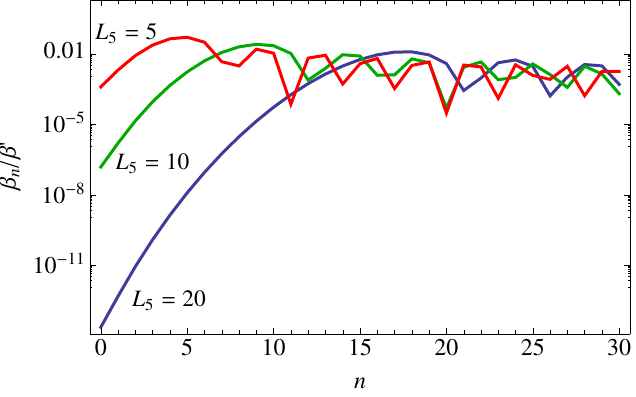}
\hspace{1mm}
\includegraphics[height=45mm]{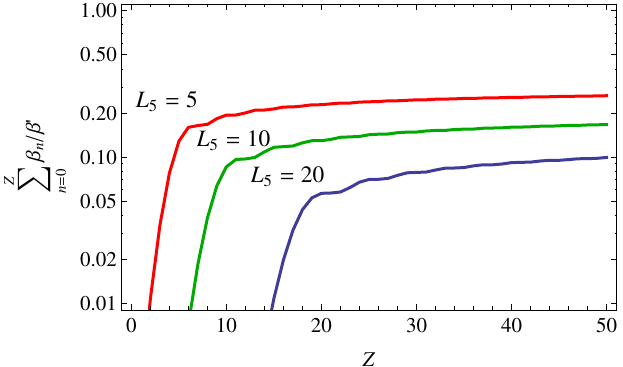}
\caption{\label{figz} 
Proton decay suppression factor $\beta_n$ for the `best' case, in which the ground state is maximally separated $x_0=\frac{L_5}{2N}$, normalised to the `worst' case $\beta'$ in which the ground state is localised on the matter curve. For different values of the compact dimension $L_5=1$ (black), 5 (red), 10 (green), 20 (blue - RHS only) and taking $N=1$ and $L_6=1$. ({\em left.}) We plot the individual suppression of each Landau level (up to $n=30$).   ({\em right.}) We show the cumulative contribution from successive modes up to $n\leq Z$ to the total suppression factor $\sum\beta_n$ relevant for the proton decay rate. 
}
\end{center}
\end{figure}

To obtain a better understanding we can examine numerical estimates of the suppression factor. Analogously to eq.~(\ref{eq5}) we shall consider ratios of suppression factors $\beta$, so as to avoid the problem of evaluating the wavefunction normalisation over the compact surface $A$. To assess the impact of localising the ground state away from the matter curve we choose to normalise $\beta_n$ relative to the `suppression' factor $\beta'$ associated to a ground state wavefunction localised on the matter curve
\beq
\beta' =
\frac{\mathcal{V}_A}{\int_A |\Phi_0|^2}\cdot \frac{1}{\pi^{1/2}d}~,
\eeq
and therefore 
\beq
\frac{\beta_n}{\beta'}\simeq
\frac{1}{2n+1}\cdot \frac{1}{2^n n!}\left[H_n\left(\frac{x_0}{d}\right)\right]^2\exp\left(-\frac{x_0^2}{d^2}\right)~.
\eeq
The left hand panel of Fig.~\ref{figz} shows the suppression factor associated to individual Landau levels $\Phi_n$, normalised to this `worst case scenario' $\beta'$ (from the point of view of extending the proton lifetime). The value of $n$ for which the ratio is maximised corresponds to the dominant state which mediates proton decay. This is typically the wavefunction which has its maximum closest to the matter curve. The level that dominates the proton decay rate is $n\approx\hat n$, as determined in eq.~(\ref{hh}). Observe, in particular, that as $L_5/Nd$ increases the ground state wavefunction $\Phi_0$ becomes exponentially suppressed, but this suppression is not exhibited by the higher modes, in line with our expectations.

It is the sum of the Landau levels which is relevant to the decay rate: $\Gamma=\Gamma_{\rm 4D}\left(\sum_{n=0}^{\infty}\beta_n\right)^2$. To obtain a handle on this overall suppression we again normalise each $\beta_n$ factor with respect to $\beta'$. In the right panel of Fig.~\ref{figz} we examine how the levels cumulatively contribute to the proton decay rate by considering the quantity  $\sum_{n=0}^{Z}\beta_n/\beta'$. We curtail the sum at an integer $Z$ and examine how varying this cut-off affects the overall suppression factor. 

The value at which the sum plateaus is approximately $Z\simeq\hat n$. For sufficiently large $Z$ this provides a good approximation to the limit $Z\rightarrow\infty$, relevant for proton decay. The plot suggests that in the thin torus limit, with large $L_5$, that $\sum\beta_n/\beta'\sim0.1$ and thus (exponentially) large suppressions can not be established. Moreover, in considering this ratio $\beta_n/\beta'$, we have omitted an explicit factor of the volume which occurs in the decay rate $\Lambda$ and, as we shall argue shortly, this volume enhancement counters the mild $L_5$ suppression observed in Fig.~\ref{figz} (right).


\subsection{Estimating the proton decay rate analytically}

To get simple, analytic estimates of the proton decay rate relative to the expectation from 4d GUTs we focus on the scenario in which the dominant decay channel is via exchange of a Landau level $\Phi_n$ with $n\gg1$ or, equivalently, $m_n^2\gg H$. This will provide us with reliable estimates if the distance to the matter curve is much larger than the width of the ground state Gaussian, $L_5/N\gg d$. The reason is that, in this situation, one clearly has to go to $n\gg 1$ before a substantial value of $\Phi_n(\frac{L_5}{2N})$ is obtained. 

By the analogy to the quantum harmonic oscillator, $|\Phi_n(x)|^2$ describes the probability of finding the highly excited particle in the quadratic potential at position $x$. For high $n$, classical intuition is good and we can express this probability parametrically in terms of the velocity, or kinetic energy, of the particle. More specifically, the probability of finding a particle, bouncing back and forth in the potential well, at the position $x$ is obviously $\sim \frac{1}{v(x)} \sim\frac{1}{\sqrt{T(x)}}$. The kinetic energy of this state is $T\sim m_n^2-H^2 x^2$ and hence and one might na\"ively want to suggest the relation
\beq
|\Phi_n(x)|^2 \sim \frac{1}{\sqrt{m_n^2-H^2 x^2}}~.\label{noave}
\eeq
However, this is taking the classical-quantum correspondence too far. The wavefunction $\Phi_n$ strongly oscillates and we can only expect the above expression to hold if we average over an interval containing at least one oscillation:
\beq 
\int_{x-\epsilon}^{x+\epsilon} {\rm d}y\, |\Phi_n(y)|^2 \sim \int_{x-\epsilon}^{x+\epsilon} \frac{{\rm d}y}{\sqrt{m_n^2-H^2 y^2}}~.\label{ave}
\eeq
Actually, all we want from this relation is an estimate of the height $\Phi_n^{\rm max}$ of the last (rightmost) peak before $\Phi_n(x)$ becomes exponentially suppressed at $x>x_{\rm max}$. Here $x_{\rm max}\equiv m_n/H$ is the maximal excursion of the classical particle with energy corresponding to the $n$th quantum state. The obvious way to obtain this estimate appears to be to `step back' from the singular point $x_{\rm max}$ by about a wave-length $\lambda$ of the oscillating function $\Phi_n$, which implies
\beq
|\Phi_n^{\rm max}|^2 
\sim 
\frac{1}{\lambda}\int^{x_{\rm max}}_{x_{\rm max}-\lambda} \frac{{\rm d}y}{\sqrt{m_n^2-H^2 y^2}} \sim \frac{1}{\sqrt{m_n^2-H^2\lambda^2}}\,.
\eeq

Thus, we now need the wavelength $\lambda$ or wave number $k=2\pi/\lambda$ of $\Phi_n$. This can be estimated by considering
\beq 
\partial^2\Phi_n=(m_n^2-H^2x^2)\Phi_n\equiv k^2\Phi_n~,
\eeq
which gives 
\beq
\lambda=\lambda(x)=\frac{2\pi}{\sqrt{m_n-H^2x^2}}~.\label{llx}
\eeq
While this diverges at $x_{\rm max}$, all we need for our estimate is the order of magnitude of the width of the last oscillation, which we call $\lambda_{\rm max}$. For this, we set $\lambda(x)=\lambda_{\rm max}$ and $x=x_{\rm max}-\lambda_{\rm max}$ in eq.~(\ref{llx}),
\beq
\lambda_{\rm max} \sim \frac{2\pi}{\sqrt{m_n^2-H^2(x_{\rm max}-\lambda_{\rm max})^2}}
\simeq
\frac{2\pi}{\sqrt{2H^2x_{\rm max}\lambda_{\rm max}}}~,
\eeq
and self-consistently solve for $\lambda_{\rm max}$. Using the approximate form of eq.~(\ref{ll}) valid for high modes,  $m_n^2\simeq2nH$, as well as $x_{\rm max}
=m_n/H$, we find
\beq
\lambda_{\rm max}\sim\frac{\pi^{2/3}x_{\rm max}}{2^{1/3} n^{2/3}}\sim \frac{x_{\rm max}}{n^{2/3}}~.
\eeq

From this, we can estimate the modification factor of eq.~(\ref{eq1}). We make the approximation that $\Phi_n$ is non-zero and constant in a region of size $\lambda_{\rm max}$ and zero otherwise. This non-zero region corresponds to the outermost peak\footnote{
We 
underestimate the proton decay rate by neglecting the contributions of the other, smaller peaks. Since we are only after a lower bound, this is acceptable.
} 
of the squared wavefunction, compare with Fig.~\ref{left}. It follows that  
\beq
\beta_n
& = |\Phi_n(x_0)|^2 \left(\int_A |\Phi_n|^2\right)^{-1}\mathcal{V}_A\cdot\frac{1}{2n+1}\\
&\sim |\Phi_n(x_0)|^2\left(\lambda_{\rm max}|\Phi_n(x_0)|^2\right)^{-1} L_5\cdot\frac{1}{2n+1}\\
&\sim \frac{L_5}{\lambda_{\rm max}}\cdot\frac{1}{2n+1}~.
\eeq
Given this, we can now evaluate eq.~(\ref{e.b}) to obtain
\beq
\frac{\Gamma}{\Gamma_{\rm 4D}}
 \sim
\left(\sum_{n=0}^\infty\frac{1}{\left(2n+1\right)}\frac{L_5}{\lambda_{\rm max}(n)}\right)^2 
 \sim
\left(\sum_{n=0}^\infty\frac{L_5}{2n^{1/3}x_{\rm max}(n)}\right)^2~,
\label{yy}
\eeq
where in the last expression we have assumed that $n\gg1$. Now we sum over $n$, subject to the constraint that the non-zero region of $\Phi_n$ contains the intersection with the matter curve. As previously, we shall assume that the matter curve is at $x_0=\frac{L_5}{2N}$. 
Let us denote the relevant range in $n$ by $[\hat{n},\hat{n}+\Delta n]$, such that 
\beq
\frac{\Gamma}{\Gamma_{\rm 4D}}\sim \left(\sum^{n=\hat n+\Delta n}_{n=\hat n}\frac{L_5}{2n^{1/3}x_{\rm max}(n)}\right)^2
\sim
\left(\frac{\Delta n L_5}{2\hat n^{1/3}x_{\rm max}(\hat n)}\right)^2~.
\eeq
Furthermore, the values of $\hat{n}$ and  $\hat{n}+\Delta n$ are fixed by the requirements
\beq 
x_{\rm max}(\hat{n}) \sim x_0
\eeq
and 
\beq
x_{\rm max}(\hat{n}+\Delta n)-x_{\rm max}(\hat{n}) \sim \lambda_{\rm max}(\hat{n})\,.
\eeq
This simply says that $\Phi_{\hat{n}}$ should have its rightmost peak at about $x_0$ and that $\Phi_{\hat{n}+\Delta n}$ should have its rightmost peak shifted by $\lambda_{\rm max}$ with respect to $\Phi_{\hat{n}}$. An estimate of the size of $\hat n$ is given in eq.~(\ref{hh}). Moreover, given that $n=H
x_{\rm max}^2(n)/2$, we then have
\beq
\frac{\Delta n}{\hat n}\simeq 
2\frac{x_{\rm max}(\hat{n}+\Delta n)-x_{\rm max}(\hat{n})}
{x_{\rm max}(\hat{n})}\sim\frac{\lambda_{\rm max}(\hat{n})}{x_{\rm max}(\hat{n})}\sim \hat{n}^{-2/3}~.
\eeq 
Thus the decay rate is parametrically
\beq
\frac{\Gamma}{\Gamma_{\rm 4D}}\sim
\left(\frac{\Delta n L_5}{2\hat n^{1/3}x_{\rm max}(\hat n)}\right)^2
\simeq
\left(\frac{L_5}{2x_{\rm max}(\hat{n})}\right)^2\sim N^2\,.
\label{res}
\eeq
This only gives a conservative lower bound on the rate as summation over modes is cut-off. Note that although $\Gamma$ is set by local data, the 4D width depends on the gauge coupling, and thus is related to the volume of the GUT brane. Therefore it is not entirely unexpected that the ratio $\Gamma/\Gamma_{4D}$ can be expressed in terms of global quantities, such as the flux $N$. However it is interesting that this parametric estimate for the ratio may be simplified to such an extent.  We conclude that the proton decay rate can not be substantially suppressed in this scenario and, further, receives an enhancement for large flux values.

Whilst a number of simplifying assumptions were made in the preceding analysis it is expected that this toy model captures all of the important physics. Specifically, we have worked under the assumption of a toroidal compactification of the internal dimensions, but we believe the crucial observation that the successive Landau levels spread out to cover the compactification surface will hold for more complicated geometries. Furthermore, whilst out analysis has been restricted to the {\bf 10-10} proton decay operators, the conclusions should remain unchanged for the analogous {\bf 10-$\bf\overline{5}$} operators.

We have argued, with a number of strong simplifications, that it seems difficult to suppress proton decay by localizing $X,Y$ gauge bosons away from the matter curve. Our reason for focussing on this possibility is that this appears to be the most obvious and potentially strongest way of suppressing proton decay. However, the analysis of [11] is to a large part based on a different possibility: The authors consider strongly localized $X,Y$ bosons, but on top of the matter curve. Based on their analysis, which only considers the lowest mode of the $X,Y$ bosons, they find a (generation dependent) non-exponential suppression also in this less extreme case. The reason is that matter fields have non-trivial profiles which are different from the profile of the $X,Y$ modes. Clearly, our analysis does not exclude this possibility. We believe that including higher Landau levels will also affect the suppression in this situation since our basic mechanism of ``effective de-localization through inclusion of higher Landau levels" should still be at work. However, we can not be certain without performing a more detailed analysis and there is clearly the option that a certain amount of suppression will be achievable. This is an interesting and important issue to be investigated in the future.

Additionally, some suppression of the proton decay rate may be possible if the component (SM) fields of the $\bf{10}$ localise in different regions of the matter curve, leading to further suppression in the wavefunction triple overlap, see e.g.~\cite{Hamada:2012wj,Font:2008id,Krippendorf:2014xba,Callaghan:2011jj,Dudas:2010zb}. 
We can envisage two distinct manners in which the components of the GUT multiplet could be localised at separated points on the matter curve.  If the flux integrated over a given matter curve $\Sigma$ is non-zero, then the components of the SU(5) representations feel the flux differently depending on their hypercharge assignment, splitting the multiplets on $\Sigma$. Thus, for $\int_\Sigma f_Y\neq0$ one can project out parts of a given GUT representation leading to incomplete GUT multiplets in the low energy theory. This is analogous to the mechanism for double-triplet splitting via flux. Importantly, anomaly cancellation requires that the net flux restricted to the matter curves $\Sigma_i$ must satisfy the Dudas-Palti relation \cite{Dudas:2010zb,Marsano:2010sq}
\beq
\sum_{a}q_a\int_{\Sigma_{{\overline5}_a}}f_Y=\sum_{i}q_i\int_{\Sigma_{{10}_i}}f_Y,
\eeq
where $q_{a,i}$ is the hypercharge of the matter fields. In this picture each generation of SM fermions does not arise from the combination of $\boldsymbol{\overline{5}}$ and $\bf 10$ of SU(5), but rather from multiple incomplete GUT multiplets. However, this scenario loses much of the motivation for GUTs as an organising principle, as outlined in Sect.~\ref{S2}.  

An alternative possibility (which preserves the motivation of Sect.~\ref{S2}) is to attempt to localise the components of a single GUT multiplet at different points on the matter curve by allowing the flux experienced by the matter curve to be locally non-zero, but in such a manner that $\int_\Sigma f_Y=0$. In the model considered in Sect.~\ref{S3} we always assume that the matter curve is aligned parallel to the direction of the flux, such that the hypercharge flux on the curve automatically vanishes. However, for more complicated embeddings of the matter curve into the GUT brane the flux incident to the matter curve could be non-zero, whilst maintaining $\int_\Sigma f_Y=0$. In this case the components of a given multiplet will localise differently. Making this effect strong likely requires some parameter tuning. It remains to be estimated how large a separation of the SM multiplets within one curve can be reasonably obtained.
Moreover, in both cases it should be shown that such constructions can be consistently realised in a global setting. 
It would also be interesting to examine whether a U(1)${}_Y$ Wilson line might allow for different localization of quarks and leptons without a splitting of the GUT multiplets. In our setting continuous Wilson lines are disfavoured as they correspond to light adjoint matter. However a localisation effect could potentially also be realised by a discrete Wilson line. For some relevant discussions see e.g.~\cite{Hamada:2012wj,Marsano:2012yc}. 
Such more complicated scenarios require dedicated studies and are beyond the scope of this paper.


\section{Limits and signals from proton decay}
\label{S4}

Given our discussions in the preceding section, henceforth, we shall assume that the overall proton decay rate is not suppressed and, consequently, an important model independent limit on gauge coupling unification is the non-observation of proton decay. Interestingly, we find that in this case, in order to satisfy the bounds on proton decay, models of F-SU(5) High Scale SUSY are required to have sub-100 TeV superpartners.
We remind the reader that our assumption of ``no suppression'' is only the most extreme case and that, if a certain amount of suppression can be realized in one of the ways discussed in the end of the last section, the upper bound on the SUSY scale will go up correspondingly.

\subsection{Proton decay constraints on dimension six operators}

\begin{figure}[t!]
\begin{center}
\includegraphics[height=60mm]{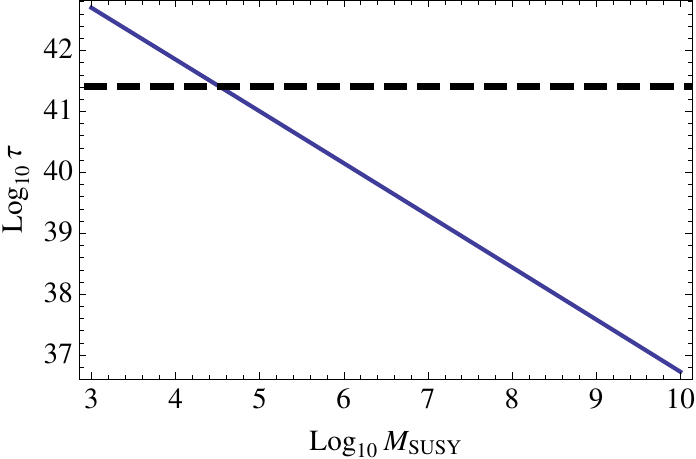}
\caption{\label{fig3}  Proton lifetime (seconds) due to decay via $X,Y$ exchange (dimension six operator) for F-SU(5) High Scale SUSY as a function of the SUSY scale (GeV). We have assumed local tadpole cancellation cf.~eq.~(\ref{333}). The dotted line indicates the experimental lower limit on $\tau(p\rightarrow\pi^0e^+)$. To satisfy this limit requires superpartner masses lighter than $M_{\rm SUSY}\lesssim30$ TeV. $M_{\rm GUT}$ and $\alpha_{\rm GUT}$ depend on the SUSY scale (via running of the gauge couplings), as given in eq.~(\ref{aa}) \& (\ref{gg}).}
\end{center}
\end{figure}

The experimental lower bound on the proton lifetime, in particular the limit on the decay channel $\tau(p\rightarrow\pi^0e^+)>2.5\times10^{41}$ s \cite{PDG}, constrains the combination of $M_{\rm GUT}$ and $\alpha_{\rm GUT}$ due to proton decay via the dimension six operator induced by $X,Y$ boson exchange. 
The corresponding decay width is given by \cite{Hisano:1992jj,Hebecker:2002rc} 
\beq
\Gamma^{X,Y}_{p\rightarrow \pi^0e^+}
&=\mathcal{C}\times\left(\frac{\alpha_{\rm GUT}^2m_p^5}{M_{\rm GUT}^4}\right)~,
\eeq
where $\mathcal{C}$ is the hadronic factor previously encountered in eq.~(\ref{e.b}). This is defined by
\beq
\mathcal{C}
=
\frac{5\pi \alpha^2}{4 f_\pi^2 m_p^4}~\left(1+D+F\right)^2A_R^2\simeq2.16~,
\eeq
where $f_\pi\simeq0.13$ GeV is the pion decay constant, $A_R\simeq2.5$ is a renormalisation constant and the parameters $D\simeq0.8$, $F\simeq0.47$ and $\alpha\simeq0.015~{\rm GeV}^3$ are determined from chiral perturbation theory
\cite{Aoki:1999tw,Aoki:2008ku}.  We can compare the parametric form of the proton decay rate due to exchange of $X,Y$ gauge bosons to the experimental limit
\beq
\tau^{X,Y}_{p\rightarrow \pi^0e^+}
\simeq 0.46\times\frac{M_{\rm GUT}^4}{\alpha_{\rm GUT}^2 m_p^5}
~>~2.5\times10^{41}~{\rm s}~.
\eeq
The proton lifetime due to $X,Y$ bosons exchange depends on $M_{\rm GUT}$ and $\alpha_{\rm GUT}$. For the F-SU(5) High Scale SUSY scenario these quantities are determined by eq.~(\ref{aa}) \& (\ref{gg})
\begin{equation*}
\begin{aligned}
M_{\rm GUT} 
&
= 4.25\times10^{15}~{\rm GeV}
\left(\frac{10^{5}~{\rm GeV}}{M_{\rm SUSY}}\right)^{2/9}~,\\[5pt]
\alpha_{\rm GUT}^{-1} 
&
\simeq28-
\frac{1}{2\pi}\frac{10}{3}
\log\left[
\left(\frac{M_{\rm SUSY}}{10^5~{\rm GeV}}\right)\right]~,
\end{aligned}
\end{equation*}
where we have assumed local tadpole cancellation as in eq.~(\ref{1.2}).
It follows that
\beq
\tau^{X,Y}_{p\rightarrow \pi^0e^+}
&\simeq \frac{0.46}{m_p^5}
\left(4.25\times10^{15}~{\rm GeV}
\left(\frac{10^{5}~{\rm GeV}}{M_{\rm SUSY}}\right)^{2/9}\right)^4
\left(28+
\frac{10}{6\pi}
\log\left[
\left(\frac{M_{\rm SUSY}}{10^5~{\rm GeV}}\right)\right]\right)^2 
\\
&\simeq 3\times10^{41}~{\rm s}~
\left(\frac{30~{\rm TeV}}{M_{\rm SUSY}}\right)^{8/9}
\left(1+
\frac{10}{168\pi}
\log\left[
\left(\frac{M_{\rm SUSY}}{30~{\rm TeV}}\right)\right]\right)^2 ~.
\label{333}
\eeq
As can be read from the equation above, and as indicated in Fig.~\ref{fig3}, the dimension six proton decay bounds are satisfied for superpartners with masses lighter than $M_{\rm SUSY}\lesssim30$ TeV. This is because as the superpartner masses are increased, the GUT scale falls resulting in the growth of the proton decay rate (slightly offset by a logarithmic increase in $\alpha_{\rm GUT}^{-1}$).  As argued in the previous section, suppression is challenging to obtain and thus spectra with high $M_{\rm SUSY}$ are disfavoured in these F-SU(5) models.

 In the case with relatively light superpartners, even in the absence of UV corrections the quality of unification is good.  Consider, for instance, a spectrum in which the superpartners lie at $30$ TeV, then the GUT scale values of  the SM gauge couplings are: 
\beq
\alpha_i^{-1}(M_{\rm GUT})=\alpha_{3}^{-1}+\delta^{\rm tree}_{i}\simeq(27.3,~28.3,~27.9)~.
\eeq 
Recall that the dominant UV correction is of the form $\delta^{\rm tree}_{i}\simeq\frac{\gamma}{g_s} b_i^{H}$. 
Comparing to the eqs.~(\ref{bh}) \& (\ref{ga}), and taking $\gamma\simeq-\frac{1}{2}\int f_Y\wedge f_Y=1$, the size of the string coupling required to account for this percent-level discrepancy between the gauge couplings is $g_s\simeq 1$.


\subsection{F-theory unification in High Scale SUSY with triplet Higgses}
\label{trip}

Doublet-triplet splitting is theoretically challenging and an intriguing possibility in High Scale SUSY models is to have the triplet Higgses at the sfermion scale. This possibility was commented upon in \cite{Ibanez:2012zg,Hisano:2013exa,Hisano:2013cqa}. However, here we shall argue that in  F-SU(5) High Scale SUSY this can not be realised without substantial suppression of the $X,Y$ boson couplings to first generation.

The presence of triplet Higgses $T$, in the representation $(\overline{3},1)_{1/3}+{\rm h.c}$., at the  scale $M_{\rm SUSY}$ implies that the UV $\beta$-function coefficients, are $b_i^{\rm UV}=\left(7, 1,-2\right)$, rather than the MSSM coefficients. However, the combination of  $\beta$-function coefficients which enters into the determination of the GUT scale $B^{\rm UV}$ has the same value as found for just the MSSM spectrum: $B^{\rm UV}=B^{\rm MSSM}$ (cf.~eq.~(\ref{B})).  Thus the value of the GUT scale is still given by eq.~(\ref{aa}).
But this does alter the GUT coupling, which now has the form
\beq
\alpha_{\rm GUT}^{-1}
&=24
+\frac{41}{18\pi}\log\left(\frac{M_{\rm SUSY}}{10^5~{\rm GeV}}\right)~.
\eeq
This leads to a mild variation in the proton lifetime due to dimension six proton decay. 
For the case that $m_T\sim M_{\rm SUSY}$ the partial proton lifetime due to the dimension six operators involving $X,Y$ gauge bosons is given by
\beq
\tau^{X,Y}_{p\rightarrow \pi^0e^+}
\simeq 2.7\times10^{41}~{\rm s}~
\left(\frac{30~{\rm TeV}}{M_{\rm SUSY}}\right)^{8/9}
\left(1+
\frac{41}{432\pi}
\log\left[
\left(\frac{M_{\rm SUSY}}{30~{\rm TeV}}\right)\right]\right)^2 ~.
\eeq
More significantly, the triplet Higgs states induce additional proton decay operators which are also strongly constrained, as we discuss next.


\subsection{Multiple proton decay signals as a signature of High Scale SUSY}

\begin{figure}[t!]
\begin{center}
\includegraphics[height=90mm]{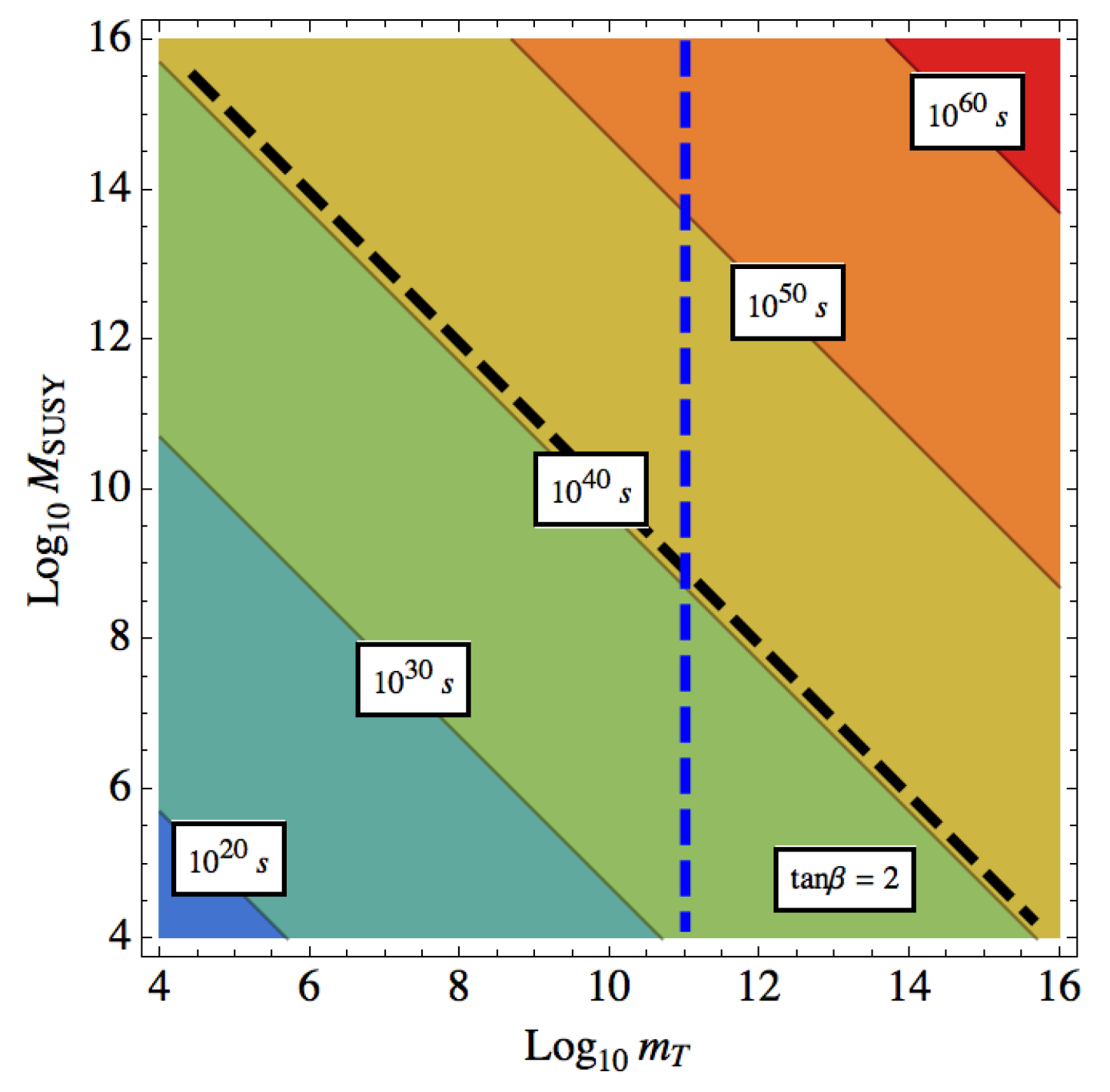}
\caption{\label{lifetime2} Contour plot showing proton lifetime in High Scale SUSY as the SUSY scale (GeV) and the scalar triplet Higgs mass  (GeV) are varied, for $\tan\beta=2$, cf.~eq.~(\ref{pp}). 
The (diagonal) black dotted line shows the lower limit on $\tau(p\rightarrow K^+\overline{\nu})$ and the (vertical) blue line indicates the lower limit from $p\rightarrow\pi^0e^+$ via triplet Higgs exchange, as described in eq.~(\ref{222}). Note that limit on $p\rightarrow\pi^0e^+$ via $X,Y$ boson exchange  favours sub-100 TeV superpartners, in which case $m_T$ must be large in order to avoid observable proton decay to $K^+\overline{\nu}$.}
\end{center}
\end{figure}

If the triplet Higgs states survive below the GUT scale then this can lead to rapid proton decay via dimension five and six effective operators, the two main decay modes which constrain this scenario are $p\rightarrow\pi^0e^+$ and $p\rightarrow K^+\overline{\nu}$ (see e.g.~\cite{Hisano:1992jj}). Similar to $X,Y$ boson exchange there is a limit on the triplet Higgs mass from $\tau(p\rightarrow\pi^0e^+)>2.5\times10^{41}$ s \cite{PDG}. The proton partial lifetime due to the dimension six operator involving the scalar triplet Higgs is parametrically (see e.g.~\cite{Nath:2006ut})  
\beq
\tau_{p\rightarrow\pi^0e^+}^{T}\sim \frac{m_T^4}{Y_d^2 Y_u^2 m_p^5}\simeq
3\times10^{41}~{\rm s}
\left(\frac{m_{T}}{10^{11}~{\rm GeV}}\right)^4
~>~2.5\times10^{41}~{\rm s}~,
\label{222}
\eeq
where $Y_f$ are the SM Yukawa couplings. 

A second important bound comes from the decay channel $\tau(p\rightarrow K^+\overline{\nu})\gtrsim2\times10^{40}$ s \cite{PDG}. This constrains the dimension five operator obtained from integrating out the fermion part of the triplet Higgs. 
The proton lifetime due to the dimension five proton decay operator is given by \cite{Hisano:2013exa}  (see also \cite{Weinberg:1981wj}),\footnote{Ref.~\cite{Hisano:2013exa} also includes some comments about the effects of renormalisation of the effective operator on the lifetime, however we neglect this sub-leading effect here, taking $\overline{A}_R=0.1$.} 
\beq
\tau_{p\rightarrow K^+\overline{\nu}}^{T}\simeq 1.2\times10^{41}~{\rm s}~\sin^42\beta
\left(\frac{M_{\rm SUSY}}{10^5~{\rm GeV}}\right)^2\left(\frac{m_T}{10^{15}~{\rm GeV}}\right)^2
\gtrsim 2\times10^{40}~{\rm s}~.
\label{pp}
\eeq
In the MSSM the factor $\sin^42\beta$ in eq.~(\ref{pp}) is fixed by the requirement that the correct Higgs mass is obtained. This is because the UV value of the quartic coupling is typically determined by the condition $\lambda=(g^2+g^\prime{}^2)\cos^22\beta/8$ and $m_H\simeq125$ GeV is found for $\tan\beta\simeq3\pm1$, see e.g.~\cite{Giudice:2011cg}. 
We observe that the relatively light superpartners ($<100$ TeV), required to avoid dimension six proton decay, lower the proton partial lifetime for the dimension five decay operator to charged kaons. Thus, in order to satisfy the bounds from current searches, the triplet Higgses must be relatively heavy $\gtrsim10^{15}$ GeV.

 Furthermore, just as we have argued here that proton decay mediated by $X,Y$ gauge bosons can not typically be suppressed in the simplest constructions, it was shown in \cite{Camara:2011nj} that the triplet-SM couplings are generally comparable or enhanced in F-theory models. Consequently, we conclude that F-SU(5) High Scale SUSY with $m_T\sim M_{\rm SUSY}$ is difficult to realise and doublet-triplet splitting near the GUT scale is likely required. Fortunately, the presence of hypercharge flux, in addition to breaking SU(5), also provides a well motivated and viable mechanism to achieve doublet-triplet splitting \cite{Mayrhofer:2013ara, Donagi:2008kj2}.

Thus there are multiple proton decay channels which are constrained by current searches and which could potentially provide complementary insights into GUT scale physics. For the High Scale SUSY spectrum with sub-100 TeV sparticles, supplemented with scalar triplet Higgses below the GUT scale, in order to avoid proton decays to $K^+\overline{\nu}$ due to the dimension five operator the triplet Higgses must be close to $M_{\rm GUT}$. In this case the partial proton lifetime due to the dimension six operator induced by triplet Higgs exchange is parametrically
  \beq
  \tau_{p\rightarrow\pi^0e^+}^{T}\simeq
3\times10^{57}~{\rm s}
\left(\frac{m_{T}}{10^{15}~{\rm GeV}}\right)^4~,
  \eeq
and therefore we should not expect signals from this channel. On the other hand, it is quite conceivable that dimension six proton decay due to $X,Y$ boson could be detected via observation of $p\rightarrow e^+\pi^0$ leading to a determination of $M_{\rm GUT}$ if $\tau_{p\rightarrow\pi^0e^+}^{(6)}\sim10^{42\pm1}$ s. Further, dimension five proton decay could be detected in the near future if $\tau_{p\rightarrow K^+\overline{\nu}}^{(5)}\sim10^{41\pm1}$ s,  
from which one could establish the mass of the triplet Higgses. The parameter space of these models is already quite constrained and it is expected that the forthcoming limits from Hyper-K \cite{Abe:2011ts} should exceed $3\times10^{42}$ s  and $6\times10^{41}$ s for $p\rightarrow e^+\pi^0$ and  $p\rightarrow K^+\overline{\nu}$, respectively, thus in principle decisively testing this class of models.


\section{Concluding remarks}
\label{Con}

Despite the apparent absence of weak scale SUSY, we have argued that GUTs still provide a strong guiding principle for BSM physics. We have attempted to clarify the precise form of the F-theory threshold corrections and discussed the prospect of F-theory gauge unification in models of High Scale SUSY, expanding upon the ideas of Ibanez {\em et al}.~\cite{Ibanez:2012zg}.  We derived general formulae for the GUT scale and GUT coupling in High Scale SUSY (along with other models of interest in the Appendices). Subsequently, we examined the limits from null searches for proton decay and the possibility of multiple channels with comparable lifetimes in High Scale SUSY. 
It was argued that, in the absence of proton decay suppression, in order to avoid observable proton decay in this setting the superpartners should have sub-100 TeV masses, as indicated in Figs.~2 \& 6. We note three caveats to this statement: First, while we have made a clear argument against exponential suppression due to $X,Y$ localization, a certain amount of suppression due to various localizations effects may be achievable. Second, higher SUSY scales can be accommodated if the MSSM spectrum is supplemented by certain additional states. An example of this with Dirac gauginos is discussed in Appendix \ref{ApA3} (see also \cite{Hall:2013eko}). Third, if there were additional large UV threshold effects from other sources, then in principle this may perturb our results.

Our conclusions are based on the crucial observation that one must include the full spectrum of higher modes when calculating the proton lifetime due to exchange of $X,Y$ gauge bosons. We have quantified to what degree the proton decay rate can be suppressed in F-theory GUTs due to localising the $X,Y$ gauge bosons away from the SM matter curves. We have argued using a 1d toy model that a suppression typically can not be achieved in the simplest scenarios, as higher modes which do not experience the exponential suppression of the ground state will dominate the proton decay rate. Any suppression due to the larger masses of these modes is compensated by their large number and their localisation. Thus, extending the proton lifetime in this manner seems difficult. 
Moreover, as remarked in the close of Sect.~\ref{S3b}, substantial suppressions of the proton decay rate may be possible if the component (SM) fields of the $\bf{10}$ localise in different  regions of the matter curve, potentially leading to further suppression in the wavefunction triple overlap, see e.g.~\cite{Hamada:2012wj,Font:2008id,Krippendorf:2014xba,Callaghan:2011jj,Dudas:2010zb}.

Even in the case that substantial suppression of proton decay can be achieved, this is probably not realized in generic models. Hence, while higher SUSY breaking would then be possible, we believe that the scenario of F-SU(5) SUSY with 100 TeV superpartners, as advocated here (building on the considerations of \cite{Ibanez:2012zg}), is an interesting and motivated framework for physics beyond the SM.
 Interestingly there have been several feasibility studies completed recently on the possibility of observing the effects of such sub-PeV SUSY partners in flavour and EDM experiments \cite{Moroi:2013sfa,McKeen:2013dma,Fuyuto:2013gla,Altmannshofer:2013lfa}. These experiments would be complementary to the proton decay searches discussed earlier. Furthermore, in this class of models at least part of the superpartner spectrum should be accessible to direct searches at a future - 100 TeV - collider  (see e.g.~\cite{Jung:2013zya}).


\section*{Acknowledgements}

We are grateful to Matthew Dolan, Emilian Dudas, Pavel Fileviez Perez, Lawrence Hall, Jonathan Heckman, Luis Ib\'a\~nez, Alexander Knochel, John March-Russell, Eran Palti, Sakura Sch\"afer-Nameki, Laura Schaposnik, Irena Valenzuela, and Timo Weigand for useful discussions.
JU is grateful for support form a Charterhouse European Bursary from the University of Oxford and the hospitality of the Institute for Theoretical Physics, Universit\"at Heidelberg.

\appendix


\section{F-theory one-loop correction}
\label{apB}

Here we provide an exposition of the derivation of the F-theory loop correction given  in \cite{Donagi:2008kj2} and highlight an apparent difference with \cite{Dolan:2011aq}. Our starting point is the following form of the gauge coupling RGE \cite{Donagi:2008kj2} in terms of contributions from the zero-modes, up to the cut-off scale $\Lambda$
\beq
\frac{16\pi^2}{g_i^2(\mu)} = 
&~ \frac{16\pi^2}{g^2}
+ 2b_i^{\rm MSSM} 
{\rm log}\left(\frac{\Lambda^2}{m_{Z}^2}\right)
+S_i+S_i'\,,
\label{b1}
\eeq
where $g$ is the unified coupling and the sums run over representations labeled by hypercharge. 
Threshold corrections due to heavy modes of the matter and gauge sector are incorporated via the terms $S$ and $S'$, respectively. The correction due to $X,Y$ modes are given by  \cite{Donagi:2008kj2}
\beq
S_i=2{\rm Tr}_{R_{5/6}}(Q_i^2)K_{5/6}+2{\rm Tr}_{R_0}(Q_i^2) K_0~,
\label{g3}
\eeq
where
\beq
2{\rm Tr}_{R_{5/6}}(Q_i^2)=
\left(
\begin{array}{c}
5\\
3\\
2
\end{array}
\right)
\qquad,
\qquad
2{\rm Tr}_{R_0}(Q_i^2)=
\left(
\begin{array}{c}
0\\
4\\
6
\end{array}
\right)~.
\eeq
The factors $K_Y$, indexed by the hypercharge $Y$, are related to the holomorphic torsion and we shall discuss the form of these quantities shortly. However, we note here that the latter term of eq.~(\ref{g3}), with $Y=0$, will not give non-universal contributions (as it is associated to a trivial line bundle with vanishing $c_1$ and ${\rm ch}_2$).  Further, the traces are related to the $\beta$-function coefficients and $2{\rm Tr}_{R_{5/6}}(Q_i^2)=b_i^{5/6}$.

The contributions of the KK modes of the matter fields $S'_i$ are given by 
\beq
S'_i= 2 \sum_{Y}{\rm Tr}_{R_Y}(Q_i^2) K_\Sigma~,
\eeq
with
\beq
K_\Sigma
&=
-H'{\rm log}\left(\frac{\Lambda^2}{M_{\rm KK}^2}\right)
-2{\bf T}(\Sigma,V)|_{R\rightarrow 1}~.
\eeq
This corresponds to a log-divergent part and an unimportant finite term, with no $\Lambda$ dependance. Henceforth we shall drop all finite and universal pieces and retain only the relevant log-divergent contributions.

As stated in \cite{Donagi:2008kj2}, the contributions of the massless modes, which are proportional to the MSSM $\beta$-function, can be expressed in terms of geometric quantities
\beq
b_i^{\rm MSSM} &= 2\left( \sum_{Y=0,\pm\frac{5}{6}}{\rm Tr}_{R_Y}(Q_i^2) H + \sum_{Y}{\rm Tr}_{R_Y}(Q_i^2)  H'\right)~,
\label{g2}
\eeq
where
\beq
H&=\left(
-\frac{3}{2}h^0(V_Y)
+\frac{1}{2}h^1(V_Y)
+\frac{1}{2}h^2(V_Y)
\right)~,\\
H'&=\left(\frac{1}{2}h^0+\frac{1}{2}h^1-\frac{1}{3}c_1(T)\right)~.
\eeq
By separating the logarithm, eq.~(\ref{b1})  can be re-expressed such that it looks like ordinary running of the zero modes up to the KK scale accompanied by various corrections 
\beq
\frac{16\pi^2}{g_i^2(\mu)} =&~ 
 \frac{16\pi^2}{g^2}+b_i^{\rm MSSM}{\rm log}\left(\frac{M_{\rm KK}^2}{m_Z^2}\right)\\
&+ 2\left( \sum_{Y=0,\pm\frac{5}{6}}{\rm Tr}_{R_Y}(Q_i^2) H + \sum_{Y}{\rm Tr}_{R_Y}(Q_i^2)  H'\right) 
{\rm log}\left(\frac{\Lambda^2}{M_{\rm KK}^2}\right)
+S_i+S_i'~.
\label{g1}
\eeq

The correction $S_i$ can be evaluated by relating the $K_Y$ to the holomorphic torsion, as discussed in \cite{Donagi:2008kj2}, via 
\beq
K_Y=K_{-Y}=2{\bf T}(S,L^Y)|_{R\rightarrow R\Lambda}~,
\label{g5}
\eeq
with the torsion given by
\begin{align}
\label{tor}
2{\bf T}(X,V)|_{R\rightarrow R\Lambda}\sim&\Bigg(\sum_q(-1)^q(2-q)h^q(X,V)\\
&-
\int_S\left[{\rm ch}_2(V_Y)
+\frac{5}{12}c_1(V_Y)c_1(T)
+\frac{1}{24}c_1(T)^2
+\frac{1}{12}c_2(T)
\right]\Bigg){\rm log}\left(\frac{\Lambda^2}{M_{\rm KK}^2}\right).
\notag
\end{align}
Evaluating the first term of the RHS of eq.~(\ref{tor}) yields
\beq
\sum(-1)^q(2-q)h^q(X,V)
&= \left(2h^0(V)-h^1(V)\right)\\
&= \frac{1}{2}\chi(V)-H~,
\eeq
and using eq.~(\ref{g2}) and the Riemann-Roch theorem
\beq
\chi(V)=
\int_S\left[
{\rm ch}_2(V_Y)
+\frac{1}{2}c_1(V_Y)c_1(T)
+\frac{1}{12}c_1(T)^2
+\frac{1}{12}c_2(T)\right]
~,
\eeq 
one obtains
\beq
\sum(-1)^q(2-q)h^q(X,V)= \int_S\left[
\frac{1}{2}{\rm ch}_2(V_Y)
+\frac{1}{4}c_1(V_Y)c_1(T)
+\frac{1}{24}c_1(T)^2
+\frac{1}{24}c_2(T)\right]-H~.
\eeq
Substituting this expression into eq.~(\ref{tor}) and simplifying leads to
\beq
2{\bf T}(X,V)|_{R\rightarrow R\Lambda}\sim&\Bigg(
 \int_S\left[
-\frac{1}{2}{\rm ch}_2(V_Y)
-\frac{1}{6}c_1(V_Y)c_1(T)
-\frac{1}{24}c_2(T)\right]-H
\Bigg){\rm log}\left(\frac{\Lambda^2}{M_{\rm KK}^2}\right)~.
\eeq
Then substituting into eq.~(\ref{g3}), one obtains the contribution of the gauge sector KK modes
\beq
S_i
&=
\Bigg(b_i^{5/6}
 \int_S\left[
-\frac{1}{2}{\rm ch}_2(V_Y)
-\frac{1}{6}c_1(V_Y)c_1(T)
-\frac{1}{24}c_2(T)\right]
\\ &\hspace{55mm}
- 2 \sum_{Y=0,\pm\frac{5}{6}}{\rm Tr}_{R_Y}(Q_i^2) H\Bigg){\rm log}\left(\frac{\Lambda^2}{M_{\rm KK}^2}\right)+\cdots~,
\eeq
and it follows that eq.~(\ref{g1}) reduces to
\begin{align}
\label{g21}
\frac{16\pi^2}{g_i^2(\mu)} &= 
 \frac{16\pi^2}{g^2}+b_i^{\rm MSSM}{\rm log}\left(\frac{M_{\rm KK}^2}{m_Z^2}\right)
+ 2 \sum_{Y}{\rm Tr}_{R_Y}(Q_i^2)  H' 
{\rm log}\left(\frac{\Lambda^2}{M_{\rm KK}^2}\right)+
S_i'\\
&+2 \sum_{Y=0,\pm\frac{5}{6}}{\rm Tr}_{R_Y}(Q_i^2)
\int_S\left[
-\frac{1}{2}{\rm ch}_2(V_Y)
-\frac{1}{6}c_1(V_Y)c_1(T)
-\frac{1}{24}c_2(T)
\right]{\rm log}\left(\frac{\Lambda^2}{M_{\rm KK}^2}\right)+\cdots.
\notag
\end{align}
Further, the contributions  $S'_i$ corresponding to the matter field excitations, as in eq.~(\ref{g3}), are given by
\beq
S'_i= 2 \sum_{Y}{\rm Tr}_{R_Y}(Q_i^2) \left[ -2{\bf T}(\Sigma,V)|_{R\rightarrow 1}
-H'{\rm log}\left(\frac{\Lambda^2}{M_{\rm KK}^2}\right) \right]~,
\eeq
and substituting this into eq.~(\ref{g21})  yields
\begin{align}
\frac{16\pi^2}{g_i^2(\mu)} &= 
 \frac{16\pi^2}{g^2}+b_i^{\rm MSSM}{\rm log}\left(\frac{M_{\rm KK}^2}{m_Z^2}\right)\\
&+ 2 \sum_{Y=0,\pm\frac{5}{6}}{\rm Tr}_{R_Y}(Q_i^2)
\int_S\left[
-\frac{1}{2}{\rm ch}_2(V_Y)
-\frac{1}{6}c_1(V_Y)c_1(T)
-\frac{1}{24}c_2(T)
\right]{\rm log}\left(\frac{\Lambda^2}{M_{\rm KK}^2}\right)+\cdots.
\notag
\end{align}
Only terms involving $V_R$ can lead to non-universal contributions and, further, only states with $Y=5/6$ contribute, for which $V_R=L^{5/6}$. Moreover, to keep the photon massless requires $\int c_1(V_R)c_1(T)=0$ and to avoid light ($Y=5/6$) exotics requires $\int {\rm ch}_2(L^{5/6})=-1$, hence
\beq\frac{16\pi^2}{g_i^2(\mu)} =
 \frac{16\pi^2}{g^2}+b_i^{\rm MSSM}{\rm log}\left(\frac{M_{\rm KK}^2}{m_Z^2}\right)
+ b^{5/6}_i{\rm log}\left(\frac{\Lambda^2}{M_{\rm KK}^2}\right)+\cdots~,
\label{b18}
\eeq
as derived in \cite{Donagi:2008kj2}.
Comparing the above to an earlier overview of  \cite{Donagi:2008kj2} appearing in  \cite{Dolan:2011aq}, an expression equivalent to eq.~(\ref{b18}) is derived before an additional zero mode contribution $b_i^{\rm MSSM}~{\rm log}(\Lambda^2/M_{\rm KK}^2)$ is added. However, this contribution is already accounted for in eq.~(\ref{b1}) and is subsequently cancelled by the KK mode contributions in deriving eq.~(\ref{b18}). Thus it appears that  \cite{Dolan:2011aq} introduce a double counting error, which explains their different result.


\section{F-theory unification with alternative spectra}

\label{S32}

\begin{table}[b!]
\begin{center}
\begin{tabular}{|c|c|c|c|c|c|c|}
\hline
SU(5) repr. & SM repr. & ~$\Delta b_1$~ & ~$\Delta b_2$~ & ~$\Delta b_3$~ & 
~$\Delta B$~\\[3pt]
\hline
$\overline{5}$ & $(1,2)_{-1/2}$ &  3/10 & 1/2 &  0 & 0 \\
\hline
$-~\prime\prime~-$ & $(\bar 3,1)_{1/3}$ & 1/5 & 0 &  1/2 &  0 \\
\hline
$10$ & ~$(\bar 3,1)_{-2/3}$ & 4/5 & 0 &  1/2 &  3/5 \\
\hline
$-~\prime\prime~-$ & $(1,1)_{1}$ & 3/5 & 0 &  0 &  3/5 \\
\hline
$-~\prime\prime~-$ & $(3,2)_{1/6}$ & 1/10 & 3/2 &  1 & $-6/5$ \\
\hline
$15$ &~$(3,2)_{1/6}$ & $-~\prime\prime~-$ & $-~\prime\prime~-$ & $-~\prime\prime~-$  & $-~\prime\prime~-$ 
\\
\hline
$-~\prime\prime~-$ & $(1,3)_{1}$ & 9/5 & 2 & 0  & $3/5$ \\
\hline
$-~\prime\prime~-$ & ~~~$(6,1)_{-2/3}$ & 8/5 & 0 & 5/2  &  $3/5$ \\
\hline
$24$ & $(1,1)_{0}$ & 0 & 0 & 0  &  $0$ \\
\hline
$-~\prime\prime~-$ & $(1,3)_{0}$ & 0 & 2 & 0  & $-6/5$ \\
\hline
$-~\prime\prime~-$ & $(8,1)_{0}$ & 0 & 0 & 3  & $-6/5$ \\
\hline
$-~\prime\prime~-$ & $(\overline{3},2)_{5/6} +$ h.c. & 5 & 3 &  2 & $12/5$ \\
\hline
\end{tabular}
\caption{Contributions to $\beta$-function coefficients from adding a full supermultiplet (see also \cite{Davies:2012vu}).\label{tab1}}
\end{center}
\end{table}

In this appendix we derive a more general formula for the F-theory GUT scale similar to that derived in Sect.~\ref{S3} and apply it to some alternative high scale SUSY spectra. We do not comment on the precise details of the UV completion into F-theory. 
We generalise the analysis of gauge unification in Sect.~\ref{S3.3} to include an additional mass scale $M_{\rm int}$; the hierarchy of scales is then $m_Z<M_{\rm Int}<M_{\rm SUSY}<M_{\rm GUT}$. We shall examine the following scenarios:
\begin{itemize}
\item Split SUSY: gauginos and Higgsinos at the intermediate scale \cite{ArkaniHamed:2004fb,Giudice:2004tc,Wells:2003tf,Arvanitaki:2012ps};
\item `Simply Unnatural SUSY': gauginos alone at the intermediate scale \cite{ArkaniHamed:2012gw,ArkaniHamed:2006mb};
\item High Scale SUSY with Dirac gauginos at the intermediate scale \cite{Unwin:2012fj}. 
\end{itemize}
The changes in  $b_i$ and $B$ (cf.~eq.~(\ref{B})) due to adding matter fields are shown in Table \ref{tab1} and the gauginos will have the following effect: $\Delta b_i=(0,4/3,2)$, thus $\Delta B=-8/5$.  Note that adding a full SU(5) multiplet, or adding matter in the representations $(1,2)_{1/2}$ or $(\overline{3},1)_{1/3}$, does not alter $B$. However, even in the case that $B$ (and thus $M_{\rm GUT}$) is unaffected, changes to the $\beta$-function will alter the running of the gauge couplings and thus the value of $\alpha_{\rm GUT}$.

We now derive an expression for the GUT scale starting from the analogue of eq.~(\ref{00}) 
\beq
\alpha_i^{-1}(m_Z) = 
&\alpha_{\rm GUT}^{-1}+
\frac{1}{2\pi}b_i^{\rm MSSM} {\rm log}\left(\frac{M_{\rm GUT}}{m_Z}\right) + \delta'_i
~,\label{B12}
\eeq 
where we have simply generalised the correction term, which now takes the form
\beq
 \delta'_i= \delta_i^{\rm tree}+ \delta_i^{\rm loop}+ \delta_i^{\rm MSSM}+\delta^{\rm Int}_i+\delta^{\rm UV}_i~.
\eeq
The first three corrections are exactly those of eq.~(\ref{susy}), (\ref{bh}) \& (\ref{5/6}).
The correction $\delta_i^{\rm Int}$ describes the departure of some of the non-SM states from the scale $M_{\rm SUSY}$
\beq
\delta^{\rm Int}_i=\frac{1}{2\pi}\left(b_i^{\rm Int}-b_i^{\rm SM}\right)
{\rm log}\left(\frac{M_{\rm SUSY}}{M_{\rm Int}}\right)~.
\eeq
The other correction $\delta_i^{\rm UV}$ describes new matter content in addition to the MSSM states which enters at the UV scale $M_{\rm SUSY}$
\beq
\delta^{\rm UV}_i=\frac{1}{2\pi}\left(b_i^{\rm UV}-b_i^{\rm MSSM}\right)
{\rm log}\left(\frac{M_{\rm GUT}}{M_{\rm SUSY}}\right)~.
\eeq
It is then straightforward to derive the analogue to eq.~(\ref{1})
\beq
\frac{1}{2\pi}B^{\rm MSSM} {\rm log}\left(\frac{M_{\rm GUT}^{(0)}}{M_{\rm GUT}}\right)
=%
\sum_{i=1}^3 \rho_i
\left(\delta_i^{\rm MSSM}+\delta_i^{\rm loop}
+\delta_i^{\rm Int}+\delta_i^{\rm UV}\right)
\label{B15}
\eeq
where $\rho_i=(1,-3/5,-2/5)$. 
It follows, after some algebra, that the GUT scale in SUSY models with two scales can be expressed (cf.~eq.~(\ref{373'})) 
\beq
{\rm log}\left(\frac{M_{\rm GUT}}{m_Z}\right)
=& \frac{B^{\rm MSSM}}{B^{\rm UV}}{\rm log}\left(\frac{M_{\rm GUT}^{(0)}}{m_Z}\right)
+\left(\frac{B^{\rm Int}-B^{\rm SM}}{B^{\rm UV}}\right)
{\rm log}\left(\frac{M_{\rm Int}}{m_Z}\right)\\
& + \left(\frac{B^{\rm UV}-B^{\rm Int}}{B^{\rm UV}}\right)
{\rm log}\left(\frac{M_{\rm SUSY}}{m_Z}\right)
-\frac{12/5}{B^{\rm UV}}~{\rm log}\left(\frac{\Lambda}{M_{\rm KK}}\right)~.
\eeq
Exponentiating, this can be re-written as follows
\beq
\left(\frac{M_{\rm GUT}}{M_{\rm GUT}^{(0)}}\right)=
\left(\frac{M_{\rm GUT}^{(0)}}{m_Z}\right)^{\frac{B^{\rm MSSM}}{B^{\rm UV}}-1}
\left(\frac{M_{\rm Int}}{m_Z}\right)^{\frac{B^{\rm Int}-B^{\rm SM}}{B^{\rm UV}}}
\left(\frac{M_{\rm SUSY}}{m_Z}\right)^{\frac{B^{\rm UV}-B^{\rm Int}}{B^{\rm UV}}}
\left(\frac{M_{\rm KK}}{\Lambda}\right)^{\frac{12/5}{B^{\rm UV}}}~.
\label{111a}
\eeq 
Notably, in several cases of interest (including those we study below) $B^{\rm Int}=B^{\rm UV}$, in which case the third factor drops out. Additionally, if $B^{\rm UV}=B^{\rm MSSM}$ (for instance if the field content is that of the MSSM), then the first factor is unity. 

\subsection{Split SUSY}

We start by considering Split SUSY \cite{ArkaniHamed:2004fb,Giudice:2004tc,Wells:2003tf,Arvanitaki:2012ps} in which the gauginos and Higgsinos are present at the intermediate scale $M_{\rm Int}=m_{1/2}$ and the non-SM states are at some UV scale $M_{\rm SUSY}=m_{\widetilde{f}}$ (below the GUT scale). 
In this scenario the UV $\beta$-function coefficients have the normal MSSM values, as stated previously, and at the intermediate scale $b_i^{\rm Int}=\left(\frac{45}{10}, -\frac{7}{6},-5\right)$.
We wish to evaluate eq.~(\ref{111a}) for this model, which depends on the following combinations of the $\beta$-function coefficients
\beq
B=b_1-\frac{3}{5}b_2-\frac{2}{5}b_3~=~\left\{\begin{array}{lc} 
\frac{44}{5}\qquad &{\rm SM}\\[5pt]
\frac{36}{5}\qquad &{\rm Int}\\[5pt]
\frac{36}{5}\qquad &{\rm UV}
\end{array}
\right.~.
\eeq
Substituting the values for $B$ into eq.~(\ref{111a}), and noting $B^{\rm Int}=B^{\rm UV}$, we can express the GUT scale just in terms of the intermediate scale $m_{1/2}$
\beq
M_{\rm GUT} &=
M_{\rm GUT}^{(0)} \left(\frac{m_{Z}}{m_{1/2}}\right)^{2/9}\left(\frac{M_{\rm KK}}{\Lambda}\right)^{1/3}
\label{split}~.
\eeq
This formula is comparable in form to that for High Scale SUSY derived earlier, eq.~(\ref{aa}), with $m_{1/2}=m_{\widetilde{f}}$, i.e.~$m_{1/2}$ is identified as the scale where the full MSSM spectrum enters, rather than an intermediate scale. 
Further, the GUT coupling  can be determined  similarly to eq.~(\ref{gg})  
\beq
\alpha_{\rm GUT}^{-1}
&\simeq 28
+\frac{10}{6\pi}
\log\left[
\left(\frac{m_{1/2}}{10^4~{\rm GeV}}\right)
\left(\frac{m_{\widetilde{f}}/m_{1/2}}{100}\right)^{3/5}
\right]~.
\label{www}
\eeq
It follows that the proton lifetime, analogous to eq.~(\ref{333}), is
\beq
\tau^{X,Y}_{p\rightarrow \pi^0e^+}
&\simeq 6\times10^{41}~{\rm s}
\left(\frac{10^4~{\rm GeV}}{m_{1/2}}\right)^{8/9}
\left(1
+\frac{10}{168\pi}
\log\left[
\left(\frac{m_{1/2}}{10^4~{\rm GeV}}\right)
\left(\frac{m_{\widetilde{f}}/m_{1/2}}{100}\right)^{3/5}
\right]\right)^2 ~,
\label{www2}
\eeq
where have assumed local tadpole cancellation, taking $\Lambda/M_{\rm KK}\simeq3.3$ as in eq.~(\ref{1.2}).
 With the parameter choices indicated the spectrum features gauginos and Higgsinos at 10 TeV and the other non-SM states at $10^6$ GeV.


\subsection{Simply Unnatural SUSY}

We can make a similar calculation for the model of Simply Unnatural SUSY \cite{ArkaniHamed:2012gw,ArkaniHamed:2006mb}, in which the Higgsinos do not enter at the low threshold and are only introduced at the same scale as the other MSSM states. For the Simply Unnatural spectrum the $\beta$-function coefficients at the intermediate scale (which features only gauginos) are $b_i^{\rm Int}=(\frac{41}{10}, -\frac{11}{6},-5)$. However, the presence (or absence) of the Higgsinos do not affect the quantity $B$ (c.f.~Table \ref{tab1}), thus
\beq
B^{\rm Int}=\frac{36}{5}~.
\eeq
This is  identical to the Split SUSY case, and the GUT scale has the same parametric form as Split SUSY, eq.~(\ref{split}). Moreover, as $b_3^{\rm Int}$ is the same as in  the Split SUSY case, the GUT coupling and, therefore also the proton partial lifetime due to  $X,Y$ exchange, are identical to eq.~(\ref{www}) \& (\ref{www2}).


\subsection{High Scale SUSY with Dirac Gauginos}
\label{ApA3}

It was argued in \cite{Unwin:2012fj} that the introduction of Dirac gauginos can be advantageous for obtaining the observed Higgs mass. This is because if the electroweak gaugino masses are dominantly Dirac, then the quartic coupling vanishes in the UV independent of the value of $\tan\beta$   \cite{Fox:2002bu}.  For related work see also \cite{Hall:2013eko}. 
The possibility of F-theory unification in models of TeV scale SUSY with Dirac gauginos was examined in \cite{Davies:2012vu}. 
Assuming that the Dirac gauginos and their scalar partners
are introduced at a single intermediate scale $m_{1/2}$, the $\beta$-function coefficients at the intermediate scale are $b_i^{\rm Int}=(\frac{41}{10}, \frac{1}{6},-2)$. The UV $\beta$-function coefficients, which includes the non-SM scalars, is given by $b_i^{\rm UV}=\left(\frac{33}{5}, 3,0\right)$ and it follows that
\beq
B^{\rm UV}= B^{\rm Int}=\frac{24}{5}~.
\eeq
Therefore $B^{\rm SM}-B^{\rm Int} = 4$ and also
note that $B^{\rm UV}\neq B^{\rm MSSM}$. Using these values we find
\beq
M_{\rm GUT} &=
\left(\frac{(M_{\rm GUT}^{(0)})^3}{m_Z}\right)^{1/2}
\left(\frac{m_{Z}}{m_{1/2}}\right)^{5/6}
\left(\frac{M_{\rm KK}}{\Lambda}\right)^{1/2}\\
&\simeq
5.5\times10^{16}~{\rm GeV}~
\left(\frac{10^{10}~{\rm GeV}}{m_{1/2}}\right)^{5/6}
\left(\frac{3.3}{\Lambda/M_{\rm KK}}\right)^{1/2}~.
\eeq
The GUT coupling  is determined to be
\beq
\alpha_{\rm GUT}^{-1}
&\simeq 30
+\frac{7}{2\pi}
\log\left[
\left(\frac{m_{1/2}}{10^{10}~{\rm GeV}}\right)\left(\frac{m_{\widetilde{f}}/m_{1/2}}{100}\right)^{2/7}
\right]~,
\eeq
and the corresponding proton partial lifetime due to $X,Y$ boson exchange is
\beq
\tau^{X,Y}_{p\rightarrow \pi^0e^+}
\simeq &~
3.4\times10^{45}~{\rm s}~
\left(\frac{10^{10}~{\rm GeV}}{m_{1/2}}\right)^{10/3}
\left(\frac{3.3}{\Lambda/M_{\rm KK}}\right)^{2}
\\
&\times \left(
1
+\frac{7}{60\pi}
\log\left[
\left(\frac{m_{1/2}}{10^{10}~{\rm GeV}}\right)\left(\frac{m_{\widetilde{f}}/m_{1/2}}{100}\right)^{2/7}\right]
\right)^2.
\label{zzt}
\eeq
With the parameter choices indicated, the spectrum features Dirac gauginos at $10^{10}$ GeV and the other non-SM states at $10^{12}$ GeV. Since  $\lambda=0$ at the SUSY scale \cite{Unwin:2012fj}, this leads to a Higgs boson mass consistent with the current experimental determination \cite{Giudice:2011cg}.  Notably, this scenario allows for heavier spectra compared to the other examples studied previously, whilst avoiding observable proton decay.


\end{document}